\documentclass[preprint,12pt]{elsarticle}
\usepackage{amssymb}

\usepackage{color, fullpage}
\journal{Computers \& Fluids}
\newcommand{\bA}{\mathbf{A}}
\newcommand{\bF}{\mathbf{F}}
\newcommand{\bU}{\mathbf{U}}

 \newdefinition{rmk}{Remark}

\begin{document}

\begin{frontmatter}

\title{HYPERBOLIC MODEL FOR FREE SURFACE SHALLOW WATER FLOWS WITH EFFECTS OF DISPERSION, VORTICITY AND TOPOGRAPHY}

\author[label1,label2]{Alexander Chesnokov}
\ead{Corresponding author: chesnokov@hydro.nsc.ru}
\author[label2]{Trieu Hai Nguyen} 
\ead{trieu.science@gmail.com}

\address[label1]{Lavrentyev Institute of Hydrodynamics SB RAS, 15 Lavrentyev Ave., Novosibirsk 630090, Russia}
\address[label2]{Novosibirsk State University, 2 Pirogova Str., Novosibirsk 630090, Russia}

\begin{abstract}
We derive a hyperbolic system of equations approximating the two-layer dispersive shallow water model for shear flows recently proposed by Gavrilyuk, Liapidevskii \& Chesnokov (J. Fluid Mech., vol. 808, 2016, pp. 441--468). The use of this system for modelling the evolution of surface waves makes it possible to avoid the major numerical challenges in solving dispersive shallow water equations, which are connected with the resolution of an elliptic problem at each time instant and realization of non-reflecting conditions at the boundary of the calculation domain. It also allows one to reduce the computation time. The velocities of the characteristics of the obtained model are determined and the linear analysis is performed. Stationary solutions of the model are constructed and studied. Numerical solutions of the hyperbolic system are compared with solutions of the original dispersive model. It is shown that they almost coincide for large time intervals. The system obtained is applied to study non-stationary undular bores produced after interaction of a uniform flow with an immobile wall, non-hydrostatic shear flows over a local obstacle and the evolution of breaking solitary wave on a sloping beach.
\end{abstract}


\end{frontmatter}

\section{Introduction} 

The second-order approximation of the shallow water theory is commonly used in modelling of the interaction of nonlinear waves. There are different versions of the governing equations taking into account the influence of non-hydrostatic pressure distribution on the structure of nonlinear surface and internal waves in the long-wave approximation~\cite{Green_Naghdi_1976, Whitham, Gavr_Tesh_2001}. Alternative formulations of these dispersive models within the framework of hyperbolic systems of equations have been proposed in~\cite{L_T_2000, Liapidevskii_Gavrilova_2008, Favrie_Gavr_2017}. The effect of non-hydrostatic distributions of pressure is used in these models by using additional `internal' variables in the equations. In \cite{Grosso_2010, Mazaheri_2016} hyperbolic techniques are applied to dispersive systems in a similar way. The main advantage of such hyperbolic approximation of the dispersive equations is essential simplification of the algorithms of numerical calculation and formulation of the boundary conditions. One example is the Serre--Green--Naghdi equations describing dispersive water waves~\cite{Green_Naghdi_1976, Serre_1953, Su_Gardner_1969}. In particular, the inversion of an elliptic operator is needed at each time step when the model is numerically solved~\cite{LeM_G_H_2010, Lannes_March_2015}. As a consequence, this drastically increases the calculation time. Another important numerical problem is how to impose artificial non-reflecting (transparent) conditions at the boundary of the calculation interval for dispersive equations~\cite{Givoli_2003, Besse_2017}. These conditions are crucial when one looks for waves passing through a bounded numerical domain. This is always an open problem for solving general dispersive equations. On the other hand, the description of full propagation of nonlinear waves from the shoaling zone to the surf zone has considerable practical importance for coastal water waves. However, dispersive models such as the Green--Naghdi equations or their hyperbolic approximation predict that no wave breaks, which is obviously incorrect. For this reason, it is needed the construction of more complex models of the theory of long waves taking into account both dispersion effects and the possibility of wave breaking. 

Shallow water flows often exhibit turbulent structures such as coastal water waves in the surf zone and rollers of hydraulic jumps. The appearance of these structures is usually related to the mechanism of wave breaking. Several experimental and theoretical works have considered the flow of hydraulic jumps or spilling breakers and studied the mechanism of turbulence generation~\cite{Svendsen_2000, Teshukov_2007, Misra_2008}. In particular, work~\cite{Misra_2008} highlighted  the importance of an intense shear layer which spreads downstream from the toe of the breaker below the spilling breaker and the recirculating region of the roller where the turbulent kinetic energy is the most intense. The development of the method for constructing depth-averaged shallow water equations for shear flows in hydrostatic approximation~\cite{Teshukov_2007} made it possible to describe flows exhibiting vortex structures such as hydraulic jump rollers~\cite{Richard_2013} and turbulent roll waves~\cite{Richard_2012, Ivanova_2017}. Numerical modelling of multi-dimensional turbulent hydraulic jumps formed in convergent radial flow was recently performed and qualitatively compared to the experimental observations in~\cite{Ivanova_2018}. However, these shallow water models for shear flows did not take into account dispersion effects. 

The approach proposed in \cite{Bonneton_2011, Tissier_2012} is based on coupling between the dispersive Green--Naghdi equations describing long waves far from the coast and the hyperbolic Saint-Venant equations describing wave breaking near the coast. The difficulty is to understand when we replace one model by the other. The search for a `switching criterion' is not a well-defined problem even if several empirical criteria have been proposed in the literature. For instance, wave phase velocity becomes larger than the flow velocity or the wave slope attains the critical value. A unified model that is capable of describing both dispersion and breaking waves, as well as the transition between different flow regimes, was derived in \cite{G_L_Ch_2016}. This model is based on a two-layer long-wave approximation of the homogeneous Euler equations with a free surface evolving over a mild slope. The upper layer is turbulent and hydrostatic, while the lower one is almost potential and can be described by Green--Naghdi equations. The interface separating these two layers is considered as a discontinuity surface where turbulent mixing occurs. The model was validated, in particular, on the propagation of Favre waves~\cite{Favre} and shoaling of solitary waves~\cite{Hsiao}. Generalization of this model for flows of stratified fluid was proposed in \cite{G_L_Ch_2019}. Recently a one-layer dispersive model of shoaling and breaking waves with shear effects complemented by turbulent viscous dissipative terms was obtained and verified in \cite{Kazakova2019}. A hierarchy of non-hydrostatic shallow water-type models approximating the incompressible Euler and Navier-Stokes systems with free surface was presented in \cite{Bristeau_2015, Fernandez-Nieto_2018}. Suggested in \cite{Escalante_2019} a two-layer dispersive system provides an efficient and accurate approach to model the propagation of waves near coastal areas and intermediate waters. 

The aim of the present paper is to derive and study a hyperbolic approximation of the proposed in \cite{G_L_Ch_2016} two-layer dispersive model for shear shallow flows. The construction of such an approximation allows us to simplify and accelerate the numerical calculations of the propagation and breaking of nonlinear waves. In Section~2 we recall a two-layer long-wave model proposed in \cite{G_L_Ch_2016} that describes the flow of a homogeneous fluid with a free surface over mild slopes, considering the effects of dispersion and vorticity. We also recall the numerical method for solving these equations, used in~\cite{G_L_Ch_2016} and based on integrating a hyperbolic system with an additional inversion of an elliptic operator at each time step~\cite{LeM_G_H_2010}. In Section~3, we give other formulation of this model, which is suitable for applying the method of constructing a hyperbolic approximation of the considered dispersive equations. Following~\cite{L_T_2000, Liapidevskii_Gavrilova_2008}, using additional `internal' variables, we derive a hyperbolic approximation of the two-layer long-wave model. Further, we find the velocities of the characteristics of this first-order system and prove that all of them are real for a sufficiently large relaxation parameter. We also perform a linear analysis and obtain the dispersion relations for the original model and its hyperbolic approximation. In Section~4, we study stationary solutions and determine the asymptotic behaviour of the solution. Numerical examples for the propagation of Favre waves, undular and monotonic turbulent bores generated by obstacle as well as the breaking of a solitary wave are considered in Section~5. In particular, we show that the obtained hyperbolic system approximates solutions of the original dispersive model. Finally, we draw some conclusions. 

\section{A two-layer long-wave approximation of the homogeneous Euler equations}

The Euler equations for two-dimensional flows can be written as 
\begin{equation}\label{eq:Euler}
 \begin{array}{l}\displaystyle
  u_t+uu_x+wu_z+\rho^{-1}p_x=0, \\[2mm]
  \varepsilon^2\big(w_t+uw_x+ww_z\big)+\rho^{-1}p_z=-g, \quad u_x+w_z=0.
 \end{array}
\end{equation}
Here $u$ and $w$ are the velocity components in the horizontal $Ox$ and vertical $Oz$ directions, $p$ is the fluid pressure, $\rho$ is the fluid density, $g$ is the gravity acceleration in vertical direction and $\varepsilon=H_0/L_0\ll 1$ is the dimensionless long wave parameter ($H_0$ and $L_0$ are the characteristic vertical and horizontal scales). Equations~(\ref{eq:Euler}) admit the conservation of energy 
\begin{equation}\label{eq:E} 
  E_t+\big((E+\rho^{-1}p)u\big)_x+\big((E+\rho^{-1}p)w\big)_z=0,
\end{equation}
where $E=(u^2+\varepsilon^2 w^2)/2+gz$. The boundary conditions at the bottom $z=b(t,x)$ and at the free surface $z=Z(t,x)$ are
\begin{equation}\label{eq:BC}  
  b_t+ub_x-w\big|_{z=b}=0, \quad Z_t+uZ_x-w\big|_{z=Z}=0, \quad p\big|_{z=Z}=0. 
\end{equation}
Following~\cite{G_L_Ch_2016}, we assume that there is the internal boundary $z=b+h$ separating the lower layer of depth $h(t,x)$, where the flow is potential, and the upper turbulent layer. The kinematic condition at this interface is
\begin{equation}\label{eq:kin-interface} 
  (b+h)_t+u(b+h)_x-w\big|_{z=b+h}=-M, 
\end{equation}
where the right-hand side $M$ is responsible for the mixing between layers. Let us introduce the depth-average velocity in the lower and upper layers and the specific turbulent energy $q^2$ for the upper layer with respect to the formulas
\[ U=\frac{1}{h}\int_b^{b+h} u\,dz, \quad \bar{u}=\frac{1}{\eta}\int_{b+h}^Z u\,dz, \quad 
   q^2=\frac{1}{\eta}\int_{b+h}^Z (u-\bar{u})^2\,dz. \]
Here $\eta(t,x)$ is the depth of the upper turbulent layer and $Z=b+h+\eta$. 

We assume the pressure distribution in the lower layer is non-hydrostatic and the flow in this domain is almost potential. The upper layer is hydrostatic and the flow in this layer is turbulent. We use an additional variable $q\geq 0$ for averaged description of the shear flow in this layer. The interaction between the layers is taken into account through a kinematic condition (\ref{eq:kin-interface}) with the right-hand side $M=\sigma q$. We also assume that $u|_{z=b+h}=U$ (this condition provides compatibility between the averaged mass, momentum and energy equations for the lower layer). Under these assumptions the following depth-averaged model for a two-layer long-wave approximation of the homogeneous Euler equations with a free surface evolving over mild slope has been derived in \cite{G_L_Ch_2016} 
\begin{equation}\label{eq:model-GLCh}
 \begin{array}{l}\displaystyle
  h_t+ (Uh)_x=-M, \quad (Uh)_t+(U^2 h+P)_x= -MU+ g\eta h_x- ghb_x, \\[4mm]\displaystyle 
  \eta_t+(\bar{u}\eta)_x=M, \quad (\bar{u}\eta)_t+ \bigg((\bar{u}^2+q^2)\eta+\frac{g\eta^2}{2}\bigg)_x= 
  MU- g\eta h_x- g\eta b_x, \\[4mm]\displaystyle
  \bigg(\bigg(\frac{\bar{u}^2+q^2+g\eta}{2}+ g(b+h) \bigg)\eta\bigg)_t + 
  \bigg(\bigg(\frac{\bar{u}^2+3q^2}{2}+ g(b+h+\eta)\bigg)\bar{u}\eta\bigg)_x= \\[4mm]\displaystyle 
  \quad\quad\quad = \bigg(\frac{U^2}{2} +g(b+h+\eta)\bigg)M+ g\eta (b+h)_t- \frac{\sigma\kappa q^3}{2}.
 \end{array}
\end{equation}
where
\[ P=gh\eta+\frac{gh^2}{2} -\frac{\varepsilon^2 h^3}{3}(\dot{U}_x-U_x^2), \quad M=\sigma q. \]
Here and below `dot' denotes the material time derivative $\dot{f}=f_t+Uf_x$. The non-negative constants $\sigma$ and $\kappa$ are the empirical parameters responsible for the mixing and energy dissipation. According to~\cite{Townsend_1956, Bradshaw_1967, L_T_2000} these parameters are as follows $\sigma\approx 0.15$ and $\kappa \in [2,6]$. A mild slope approximation means that the dimensionless bottom variation is weak~\cite{Serre_1953, G_L_Ch_2019}: $z=b(\varepsilon^\gamma t, \varepsilon^\gamma x)$, $\gamma>0$. Due to this fact the terms $\varepsilon^2\ddot{b}$ and $\varepsilon^2\dot{b}$ can be neglected during the derivation of system~(\ref{eq:model-GLCh}).  

The first two equations in (\ref{eq:model-GLCh}) correspond to the balance of mass and momentum in the non-hydrostatic and almost potential lower layer. They are obtained by averaging the incompressibility and horizontal momentum equations in (\ref{eq:Euler}) over the depth taking into account the boundary conditions~(\ref{eq:BC}) and~(\ref{eq:kin-interface}). We also assume that $u|_{z=b+h}=U$, since this is the only case the first two equations in (\ref{eq:model-GLCh}) are compatible with the energy equation for the lower layer (see~\cite{G_L_Ch_2016} for details)
\[ \begin{array}{l}\displaystyle 
    \bigg(\bigg(\frac{U^2+gh}{2}+ gb+ \frac{\varepsilon^2}{6} h^2U_x^2\bigg)h\bigg)_t +
    \bigg(\bigg(\frac{U^2+gh}{2}+ gb+ \frac{\varepsilon^2}{6} h^2U_x^2\bigg)Uh +UP\bigg)_x = \\[4mm]\displaystyle
    \quad\quad\quad = 
    -\bigg(\frac{U^2}{2}+ g(b+h+\eta)+ \frac{\varepsilon^2}{2} h^2U_x^2\bigg) M -g\eta h_t +ghb_t.
   \end{array} \]

The averaged mass and momentum balance equations for the upper hydrostatic layer are derived similarly. However, the flow in this layer is vortex and in the averaged description it is characterized by shear velocity $q$ defined above. In order to obtain the closing relation, we average the energy equation~(\ref{eq:E}) by the upper layer depth. It was shown in \cite{G_L_Ch_2016} that it is convenient to use the differential consequence
\begin{equation}\label{eq:q} 
  q_t+(\bar{u}q)_x= \frac{\sigma}{2\eta}\big((U-\bar{u})^2-(1+\kappa)q^2\big)
\end{equation}
instead of the last balance law in (\ref{eq:model-GLCh}).

The upper turbulent layer is considered within the hydrostatic approximation for shear shallow water flows~\cite{Teshukov_2007}, while the lower layer can be described by the Serre--Green--Naghdi dispersive equations~\cite{Green_Naghdi_1976, Serre_1953}. The interaction between these layers is taken into account through a natural mixing process. This model reveals the main mechanism of the spilling breaker development. For waves of moderate amplitude the upper thin layer is dynamically passive and the flow is governed by the dispersive model. However, for the waves of larger amplitude, the upper turbulent layer dynamics becomes crucial. The larger the wave amplitude, the greater the difference is in the average flow velocities of layers. It finally results in an intensive growth of the upper turbulent layer.  

For the numerical solution of system~(\ref{eq:model-GLCh}) in \cite{G_L_Ch_2016} it was proposed to introduce a new variable $K$ as follows~\cite{LeM_G_H_2010}
\begin{equation}\label{eq:K-def}  
  K=U-\frac{\varepsilon^2}{3 h} \big(h^3 U_x\big)_x 
\end{equation}
and instead of the second equation in (\ref{eq:model-GLCh}) to use its differential consequence
\begin{equation}\label{eq:K}  
  K_t +\bigg( KU +g(h+\eta) -\frac{U^2}{2} -\varepsilon^2\bigg(\frac{U_x^2 h^2}{2}+ 
  \sigma q h U_x\bigg) \bigg)_x = -\frac{\varepsilon^2}{3}\sigma q h U_{xx} -gb_x.
\end{equation}
Then the numerical resolution of the governing equation is divided into two successive steps: the time evolution of the conservative variables $\mathbf{u}=(h, K, \eta, \bar{u}\eta, q)$ using the method based on modification of the Godunov's scheme; and the resolution of an ordinary differential equation~(\ref{eq:K-def}) to obtain the values of velocity $U$ from variables $h$ and $K$.

\section{Hyperbolic approximation of the two-layer dispersive equations for shear flows}

At the beginning of this section, we recall the method for constructing the hyperbolic approximation of a dispersive system proposed in \cite{L_T_2000, Liapidevskii_Gavrilova_2008}. Further, we apply this method to model~(\ref{eq:model-GLCh}) and derive a first-order evolutionary system that approximates this dispersive model. 
We also study the hyperbolicity of the obtained system and perform a standard Fourier analysis of linearised equations.

\subsection{Method for constructing a hyperbolic approximation of a dispersive model}

As in the example of the following one-dimensional dispersive system 
\begin{equation}\label{eq:class} 
 \begin{array}{l}\displaystyle
  \rho_t+(u\rho)_x=0, \quad (u\rho)_t+(u^2\rho+P)_x=0, \\[3mm]\displaystyle 
  P=P\bigg(\rho,\frac{d\rho}{dt},\frac{d^2\rho}{dt^2}\bigg), \quad 
  \frac{d}{dt}=\frac{\partial}{\partial t}+u\frac{\partial}{\partial x}
 \end{array} 
\end{equation}
we briefly explain the method of constructing a hyperbolic approximation of (\ref{eq:class}) on the basis of the averaging of instantaneous variables \cite{L_T_2000, Liapidevskii_Gavrilova_2008}. System~(\ref{eq:class}) includes the equations of bubbly flow, models of the shallow water theory, etc. \cite{Gavr_Tesh_2001, Green_Naghdi_1976, Whitham}. Such equations are used to describe flows with `internal inertia', i.e., heterogeneous media where a certain part of the total energy transforms to the energy of small-scale motion. Numerical implementation of dispersive models involves some difficulties caused by non-hyperbolicity of the considered system. 

A hyperbolic model is obtained by additional averaging of the equations and introducing new `internal' variables. The scale of averaging is assumed to be rather small, which allows the values of the variables $\rho$ and $u$ to be replaced in the equations by their mean values. However, to calculate the function $P$, derivatives of the `instantaneous' variable $\tilde{\rho}$ are used. This means that 
\[ P=P\bigg(\rho,\frac{d\tilde{\rho}}{dt},\frac{d^2\tilde{\rho}}{dt^2}\bigg).\] 
The relation between the averaged and `internal' variables is given by expanding the functions $\tilde{\rho}(s)$ into a Taylor series along the trajectory $x=x(s,\xi)$
\[ \tilde{\rho}(s)=\tilde{\rho}(t)+\tilde{\rho}'(t)(s-t)+\tilde{\rho}''(t)(s-t)^2/2+o(\tau^2), 
   \quad s\in(t-\tau,t+\tau). \]
Here $\xi$ is the fixed Lagrangian coordinate of the particle. It follows from the previous formula that the mean value $\rho(t)$ and instantaneous variable $\tilde{\rho}(t)$ are related by 
\[ \rho(t)=\frac{1}{2\tau} \int\limits_{t-\tau}^{t+\tau} \tilde{\rho}(s)\,ds= \tilde{\rho}(t)+ 
   \frac{1}{6}\tilde{\rho}''(t)\tau^2+ o(\tau^2) \]
and, consequently, 
\[ \tilde{\rho}''(t)= \alpha(\rho(t)-\tilde{\rho}(t)) +O(\tau), \quad \alpha=6/\tau^2. \]
Using the main part in representation of function $\tilde{\rho}''(t)$, we can approximate system~(\ref{eq:class}) as follows \cite{Liapidevskii_Gavrilova_2008}
\begin{equation}\label{eq:class-hyp}  
 \begin{array}{l}\displaystyle 
   \rho_t+(u\rho)_x=0, \quad (u\rho)_t+(u^2\rho+\bar{P})_x=0, \\[2mm]\displaystyle 
   \tilde{\rho}_t+u\tilde{\rho}_x=v, \quad v_t+uv_x=\alpha (\rho-\tilde{\rho}),
  \end{array} 
\end{equation}
where $\bar{P}=(\rho,\tilde{\rho},v)=P(\rho,v,\alpha(\rho-\tilde{\rho}))$. For a wide class of fluid flows, Eqs.~(\ref{eq:class-hyp}) form a hyperbolic system with two sonic and two contact characteristics. As the parameter $\alpha$ increases (in this case the averaging interval $\tau$ decreases), the solutions of hyperbolic Eqs.~(\ref{eq:class-hyp}) approximate the solutions of dispersive system~(\ref{eq:class}).

\subsection{Hyperbolic approximation of Eqs.~(\ref{eq:model-GLCh})}

Let us modify system (\ref{eq:model-GLCh}) so that it is convenient to apply the method of constructing a hyperbolic approximation described above. Taking into account the first equation in (\ref{eq:model-GLCh}) one can deduce that
\[ U_x=-\frac{\dot{h}+\sigma q}{h}, \quad 
   \dot{U}_x=-\frac{\ddot{h}+\sigma\dot{q}}{h}+\frac{(\dot{h}+\sigma q)\dot{h}}{h^2} \]
and, consequently, the dispersive term in the second equation of (\ref{eq:model-GLCh}) reads
\[ -\frac{\varepsilon^2 h^3}{3}\big(\dot{U}_x- U_x^2\big)= 
   \frac{\varepsilon^2 h^2}{3}\ddot{h}+O(\sigma\varepsilon^2). \]
We recall that the governing equations for the lower dispersive layer have been derived with accuracy $O(\varepsilon^2)$. Since the mixing parameter $\sigma\approx 0.15$ is also small ($\sigma \sim \varepsilon$), in this approximation the terms of order $O(\sigma\varepsilon^2)$ can be omitted. That is why with the same order of accuracy we can use the second equation of system~(\ref{eq:model-GLCh}) in the form
\begin{equation}\label{eq:Momentum_LL} 
  (Uh)_t+ \bigg(U^2 h+ \frac{gh^2}{2}+ gh\eta +\frac{\varepsilon^2 h^2}{3}\ddot{h} \bigg)_x 
  =g\eta h_x- \sigma qU -ghb_x. 
\end{equation}
This form of the momentum equation for the lower dispersive layer is more convenient for constructing a first-order approximative system. 

From system (\ref{eq:model-GLCh}) one can clearly see conservation of the total mass and momentum (for flows over a flat bottom). We note that up to the small terms of order $O(\sigma\varepsilon^2)$ the total energy also conserves if the additional dissipation term vanishes ($\kappa=0$).

Following \cite{Liapidevskii_Gavrilova_2008}, we introduce the new instantaneous variables $\zeta$ and $V$ so that $\dot{\zeta}=V$, $\dot{V}=\alpha(h-\zeta)$ and replace the term $\ddot{h}$ in (\ref{eq:Momentum_LL}) by $\alpha(h-\zeta)$. In this case, the approximation of Eqs.~(\ref{eq:model-GLCh}) by a system of the first-order evolutionary equations is written as follows
\begin{equation}\label{eq:hyp-appr}
 \begin{array}{l}\displaystyle
  (h+\eta)_t+ (Uh+\bar{u}\eta)_x=0, \quad \eta_t+(\bar{u}\eta)_x=\sigma q, \\[4mm]\displaystyle 
 (Uh)_t+ \bigg(U^2 h+ \frac{gh^2}{2}+ gh\eta +\frac{\alpha\varepsilon^2}{3}(h-\zeta)h^2\bigg)_x 
  =g\eta h_x- \sigma qU -ghb_x, \\[4mm]\displaystyle  
 (\bar{u}\eta)_t+ \bigg((\bar{u}^2+q^2)\eta+\frac{g\eta^2}{2}\bigg)_x= -g\eta h_x +\sigma qU- g\eta b_x, \\[4mm]\displaystyle 
  q_t+(\bar{u}q)_x= \frac{\sigma}{2\eta}\big((U-\bar{u})^2-(1+\kappa)q^2\big), \\[4mm]\displaystyle 
 (h\zeta)_t+(Uh\zeta)_x=Vh-\sigma q\zeta, \quad (Vh)_t+(UVh)_x=\alpha(h-\zeta)h-\sigma qV.
 \end{array}
\end{equation}
Here we represent equations for the variables $\zeta$ and $V$ in a conservative form (the last two equations in (\ref{eq:hyp-appr})) that is convenient for numerical treatment. There are non-conservative terms $\pm g\eta h_x$ on the right-hand side of momentum equations in (\ref{eq:model-GLCh}) and (\ref{eq:hyp-appr}). This is admissible because in the lower layer the flow is described by the Green--Naghdi equations (or their hyperbolic approximation) having smooth solutions. 

System (\ref{eq:hyp-appr}) contains parameter $\alpha$ depending on the chosen scale of averaging of the original model~(\ref{eq:model-GLCh}). The solutions of Eqs.~(\ref{eq:hyp-appr}) approximate the solutions of Eqs.~(\ref{eq:model-GLCh}) as $\alpha\to\infty$. They also yield the solutions of (\ref{eq:model-GLCh}) in the hydrostatic case ($\varepsilon=0$) if $\alpha\to 0$. We note that model (\ref{eq:model-GLCh}) in hydrostatic approximation was derived and studied in \cite{Liapidevskii_Chesnokov_2014}. To describe such flows we eliminate the last two equations and rewrite the third and fourth equations of (\ref{eq:hyp-appr}) in the  conservative form
\[ U_t+ \bigg(\frac{U^2}{2}+gH\bigg)_x=-gb_x, \quad 
   Q_t+\bigg(U^2 h+(\bar{u}^2+q^2)\eta+\frac{gH^2}{2}\bigg)_x=-gHb_x, \]
where $H=h+\eta$ and $Q=Uh+\bar{u}\eta$ are the total depth and flow rate respectively. 

\subsection{Characteristics of Eqs.~(\ref{eq:hyp-appr})}

Let us find the characteristics of system (\ref{eq:hyp-appr}). It follows from~(\ref{eq:hyp-appr}) that there are three contact characteristics $dx/dt=U$ (multiplicities two) and $dx/dt=\bar{u}$ since the variables $\zeta$, $V$ and $s=q/\eta$ satisfy the equations
\[ \begin{array}{l}\displaystyle 
    \zeta_t+U\zeta_x=V, \quad V_t+UV_x=\alpha(h-\zeta), \\[4mm]\displaystyle
    s_t+\bar{u}s_x=\frac{\sigma}{2\eta^2}\big((U-\bar{u})^2-(3+\kappa)q^2\big). 
   \end{array} \]
The momentum equations for the lower and upper layers in system~(\ref{eq:hyp-appr}) are equivalent to
\[ \begin{array}{l}\displaystyle 
    U_t+UU_x -\frac{\alpha h}{3}\zeta_x +    
    \bigg(g+\alpha\bigg(h-\frac{2\zeta}{3}\bigg)\bigg)h_x +g\eta_x = -gb_x, \\[4mm]\displaystyle
    \bar{u}_t + \bar{u}\bar{u}_x + 2\eta^2 ss_x + gh_x + (g+3s^2\eta)\eta_x = \sigma s(U-\bar{u}) - gb_x.
   \end{array} \]
Here and below we assume that $\varepsilon=1$. 

Using the previous equations, we represent system~(\ref{eq:hyp-appr}) in the form
\[ \bU_t+\bA\bU_x=\bF \]
where $\bU=(\zeta,V,s,h,\eta,U,\bar{u})^{\rm T}$ is the vector of unknown variables, $\bA$ is the $7\times 7$ matrix, and $\bF$ is the right-hand side. The eigenvalues of $\bA(\bU)$ are determined by equation
\[ \chi(\lambda)=(U-\lambda)^2(\bar{u}-\lambda)\hat{\chi}(\lambda)=0, \]
where
\[ \begin{array}{l}\displaystyle 
    \hat{\chi}(\lambda)= \big((U-\lambda)^2 -a_1\big) \big((\bar{u}-\lambda)^2-a_2\big)-g^2h\eta, \\[3mm]\displaystyle 
    a_1=gh+\alpha h \bigg(h-\frac{2\zeta}{3}\bigg), \quad a_2=g\eta+3q^2.
   \end{array} \]
Obviously, for a single-layer flow ($\eta=0$ or $h=0$), the system is hyperbolic and the sonic characteristics are 
\[ \lambda_{1,2}= U\pm \sqrt{a_1}, \quad \lambda_{3,4}= \bar{u}\pm \sqrt{a_2}.\]

\begin{figure}[t]
\begin{center}
\resizebox{.95\textwidth}{!}{\includegraphics{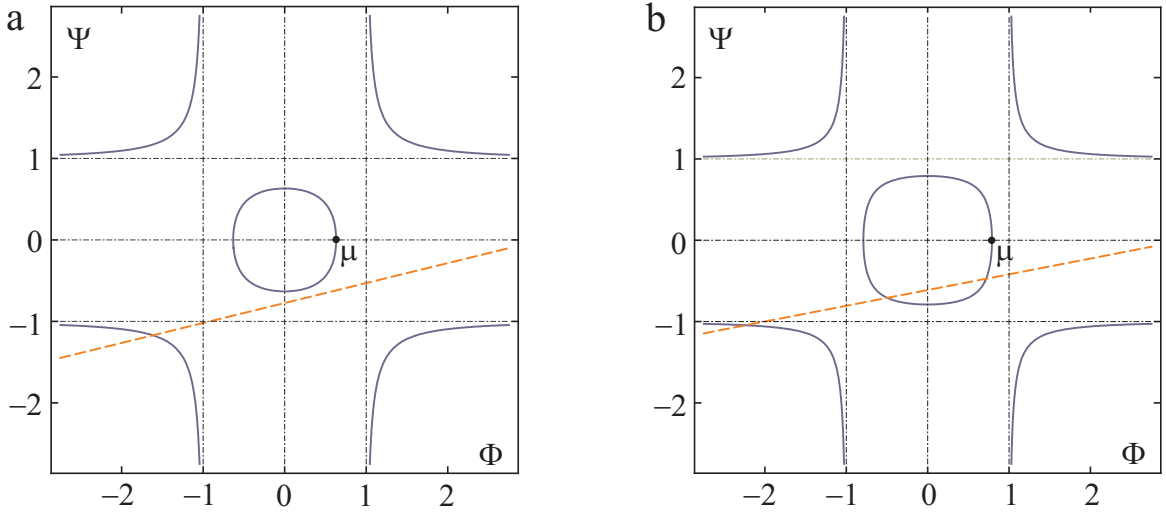}}\\[0pt]
{\caption{The curves (\ref{eq:curve}) and the straight lines (\ref{eq:line}) in the $(\Phi,\Psi)$--plane for $g=1$, $h=\zeta=1$, $\eta=0.1$, $U=0.5$, $\bar{u}=1.5$, and $q=0$ : (a) --- $\alpha=2$; (b) --- $\alpha=5$.} \label{fig:fig_char}} 
\end{center}
\end{figure}

Let us show that for a two-layer flow ($h>0$, $\eta>0$ and $h>2\zeta/3$) system (\ref{eq:hyp-appr}) is hyperbolic for a sufficiently large value of $\alpha$. A geometric interpretation of the characteristics proposed in \cite{Ovs_1979, Chesn_2017} for two-layer hydrostatic flows can be applied here. We introduce the new variables $\Phi$ and $\Psi$ by the formulas
\begin{equation}\label{eq:Phi-Psi} 
  \Phi=(\bar{u}-\lambda)/\sqrt{a_2}, \quad \Psi=(U-\lambda)/\sqrt{a_1}. 
\end{equation}  
Then equation $\hat{\chi}(\lambda)=0$ can be rewritten in the form
\begin{equation}\label{eq:curve} 
 (\Phi^2-1)(\Psi^2-1)=r,
\end{equation}  
where $r=(a_1 a_2)^{-1}g^2 h\eta$. In the $(\Phi, \Psi)$--plane, Eq.~(\ref{eq:curve}) describes a fourth-order curve with four symmetry axes (see Fig.~\ref{fig:fig_char}). The variables $\Phi$ and $\Psi$ by virtue of (\ref{eq:Phi-Psi}) are related by
\begin{equation}\label{eq:line} 
 \Psi=\Phi\sqrt{a_2/a_1}+(U-\bar{u})/\sqrt{a_1}.
\end{equation}

The number of real roots of $\hat{\chi}(\lambda)=0$ is determined by the number of intersections of the curve (\ref{eq:curve}) with the straight line (\ref{eq:line}). Each point of interaction yields a sonic characteristic with the slope $\lambda=U-\Psi\sqrt{a_1}$. It follows from Fig.~\ref{fig:fig_char} that line (\ref{eq:line}) always has two points of intersection with curve (\ref{eq:curve}) in the domains $\{\Phi>1, \Psi>1\}$ and $\{\Phi<-1, \Psi<-1\}$. We note that the necessary condition for the existence of four real roots is the inequality $\mu=\sqrt{1-r}>0$, which is always fulfilled. If we increase the parameter $\alpha$, the `radius' $\mu$ becomes larger (it tends to unity for $\alpha\to\infty$) while an initial ordinate $(U-\bar{u})/\sqrt{a_1}$ of the straight line (\ref{eq:line}) tends to zero. As a result, for sufficiently large $\alpha$ there are two points of intersection of the line with curve (\ref{eq:curve}) in the square $\{-1<\Phi<1, -1<\Psi<1\}$. 

For all numerical examples considered below, Eqs.~(\ref{eq:hyp-appr}) are hyperbolic (characteristic equation $\chi(\lambda)=0$ has seven real roots). The presence of real characteristics makes it possible to apply the standard Godunov-type schemes for numerical solving system of balance laws~(\ref{eq:hyp-appr}). We note that the increase in parameter $\alpha$ also increases the absolute value of the maximum characteristic velocity. Time step in the Godunov-type schemes is inversely proportional to the characteristic velocity. Therefore, increasing of $\alpha$ slows down the computation.

\subsection{Linear analysis}

In this section, dispersion relations of the two-layer dispersive system (\ref{eq:model-GLCh}) and its hyperbolic approximation (\ref{eq:st-hyp-appr}) are obtained and studied. This analysis allows one to follow the influence of the relaxation parameter $\alpha$ on the accuracy of the hyperbolic approximation. Here we assume a flat bottom topography $b=0$ and take $\varepsilon=1$. 

Consider the perturbation of a constant state $h=h_0$, $U_0$, $\eta=\eta_0$, $\bar{u}=U_0$, $q=0$ for dispersive system (\ref{eq:model-GLCh}). Since the governing equations are Galilean invariant, without loss of generality, we can choose $U_0=0$. At first order the system reads: 
\begin{equation}\label{eq:linear-system} 
 \begin{array}{l}\displaystyle 
   h_t+h_0U_x=-\sigma q, \quad U_t+gh_x+g\eta_x+3^{-1}h_0 h_{ttx}=0, \\[2mm]\displaystyle
   \eta_t+\eta_0\bar{u}_x=\sigma q, \quad \bar{u}_t+gh_x+g\eta_x=0, \quad q_t=0.
  \end{array} 
\end{equation}
The search for a non-trivial solution of the linear system in the form 
\[ (h, U, \eta, \bar{u}, q)=(h_1, U_1, \eta_1, \bar{u}_1, q_1)\exp(ik(x-ct)) \]
leads to the following dispersion relation 
\begin{equation}\label{eq:disp-rel-1} 
  c^2=\frac{g h_0}{1+k^2h_0^2/3}+g\eta_0.
\end{equation}
Here $k$ is the wave number and $c$ is the phase velocity. Note that for a single-layer non-hydrostatic flow ($\eta_0=0$) formula (\ref{eq:disp-rel-1}) transforms to the well-known dispersion relation of the Serre--Green--Naghdi equations. 

\begin{figure}[t]
\begin{center}
\resizebox{.6\textwidth}{!}{\includegraphics{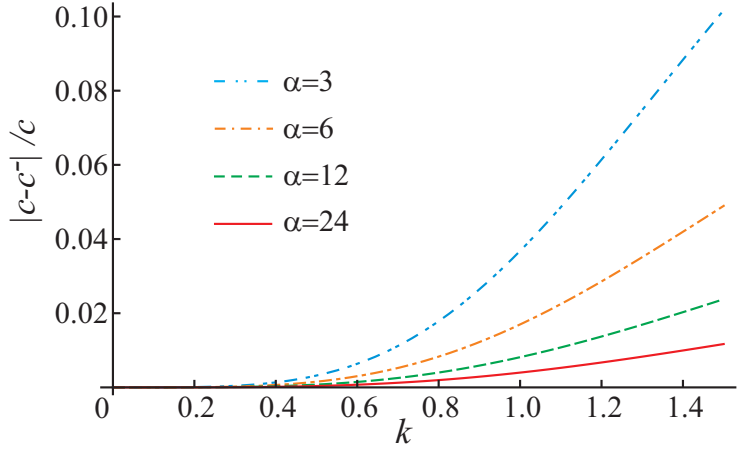}}\\[0pt]
{\caption{The relative error between the phase velocities $c$ and $c^-(\alpha)$ is shown as a function of the wave number $k$ with $g=1$, $h_0=1$, $\eta_0=0.1$.} \label{fig:fig_disp}} 
\end{center}
\end{figure}

Let us linearise hyperbolic model (\ref{eq:hyp-appr}) on the same constant solution (additionally we suppose $\zeta=h_0$ and $V=0$). Then the linear system takes form (\ref{eq:linear-system}) where the second equation reads
\[ U_t+\Big(g+\frac{\alpha h_0}{3}\Big)h_x+g\eta_x-\frac{\alpha h_0}{3}\zeta_x=0 \]
and two more equations $\zeta_t=V$ and $V_t=\alpha(h-\zeta)$ should be added to the system. As before, we are looking for a solution in the form of monochromatic perturbations (with $c\neq 0$). In this case a non-trivial solution exists if the phase velocity $c$ satisfies the equation
\[ c^4- \Big(gH_0+\frac{\alpha h_0^2}{3}+\frac{\alpha}{k^2}\Big) c^2+ 
   \alpha g\Big(\frac{H_0}{k^2} +\frac{\eta_0 h_0^2}{3}\Big) =0, \]
where $H_0=h_0+\eta_0$. The previous equation (at least for $\eta_0<h_0+(h_0^2/6+1/k^2)\alpha/g$) has two real positive roots $(c^\pm)^2$: 
\begin{equation}\label{eq:disp-rel-2} 
  (c^\pm)^2=\frac{1}{2}\Big(gH_0+\frac{\alpha h_0^2}{3}+\frac{\alpha}{k^2}\Big)\pm 
  \sqrt{\frac{1}{4}\Big(gH_0+\frac{\alpha h_0^2}{3}+\frac{\alpha}{k^2}\Big)^2 
  -\alpha g\Big(\frac{H_0}{k^2}+\frac{h_0^2\eta_0}{3}\Big) }\,.
\end{equation}
It follows from (\ref{eq:disp-rel-1}) and (\ref{eq:disp-rel-2}) that the phase velocities $c^\pm$ and $c$ satisfy to inequalities
\[ -c^+(\alpha)<-c<-c^-(\alpha)<0<c^-(\alpha)<c<c^+(\alpha), \]
and $c^-(\alpha)\to c$ as $\alpha\to\infty$ (see Fig.~\ref{fig:fig_disp}). We note that similar inequalities for the phase velocities were obtained in \cite{Favrie_Gavr_2017} for one-layer flows ($\eta_0=0$). 

The linear analysis allows us to estimate the values of the relaxation parameter $\alpha$ to achieve the required approximation accuracy using model~(\ref{eq:hyp-appr}). As we can see from Fig.~\ref{fig:fig_disp} with the given data and $\alpha=6$ the relative error $|c-c^-|/c$ between the phase velocities of the hyperbolic model and the original dispersive system is not more than 5\,\% for $k\in (0,1.5)$. With $\alpha=12$ and $\alpha=24$ accuracy increases to about 2\,\% and 1\,\%, respectively. Thus, a comparison of the dispersion relations (\ref{eq:disp-rel-1}) and (\ref{eq:disp-rel-2}) gives a criterion for choosing $\alpha $, which provides the specified accuracy of approximation between these models. 

\section{Stationary solutions}

Continuous stationary solutions of model~(\ref{eq:hyp-appr}) for flows over a flat bottom satisfy the equations
\begin{equation}\label{eq:st-hyp-appr}
 \begin{array}{l}\displaystyle
  (Uh)'= -\sigma q, \quad 
  UU'+g\eta'- \frac{\alpha}{3}h\zeta' + 
  \bigg(g+\alpha\bigg(h-\frac{2\zeta}{3}\bigg)\bigg)h' =0, \\[4mm]\displaystyle  
  (\bar{u}\eta)'= \sigma q, \quad \bar{u}\bar{u}' +2qq' +gh' +\bigg(g+\frac{q^2}{\eta}\bigg)\eta' = 
  \frac{\sigma q}{\eta}(U-\bar{u}), \\[4mm]\displaystyle  
  (\bar{u}q)'=\frac{\sigma}{2\eta}\big((U-\bar{u})^2-(1+\kappa)q^2\big), \quad 
  U\zeta'=V, \quad UV'=\alpha(h-\zeta). 
 \end{array}
\end{equation}
Here we denote the derivative with respect to $x$ by `prime' and take $\varepsilon=1$. We also point out that the governing equations~(\ref{eq:hyp-appr}) (as well as original model (\ref{eq:model-GLCh})) are Galilean invariant if $b=0$. Therefore, for flows over a flat bottom, solutions of Eqs.~(\ref{eq:hyp-appr}) in the class of travelling waves are described by equations~(\ref{eq:st-hyp-appr}). 

The normal form of Eqs.~(\ref{eq:st-hyp-appr}) is
\begin{equation} \label{eq:st-norm} 
  \begin{array}{l}\displaystyle 
    h'=\frac{G}{\Delta}, \quad \eta'=\frac{g\eta}{\varphi}(h'+b')+ 
    \frac{\sigma q\psi}{\bar{u}\varphi}, \quad \zeta'=\frac{V}{U}, \\[4mm]\displaystyle 
    V'=\frac{\alpha(h-\zeta)}{U}, \quad U'=-\frac{Uh'+\sigma q}{h}, \quad \bar{u}'=\frac{\sigma q-\bar{u}\eta'}{\eta}, \\[4mm]\displaystyle
    q'=\frac{q\eta'}{\eta}+\frac{\sigma}{2\eta\bar{u}}\big((U-\bar{u})^2-(3+\kappa)q^2\big), 
  \end{array} 
\end{equation}
where
\[ \begin{array}{l}\displaystyle  
    \varphi=\bar{u}^2-g\eta-3q^2, \quad \psi=U^2-3(U-\bar{u})\bar{u} -(3+\kappa)q^2, \\[4mm]\displaystyle 
    \Delta=1-\frac{U^2}{gh}+\frac{g\eta}{\varphi}+ \frac{\alpha}{g}\Big(h-\frac{2\zeta}{3}\Big), \quad
    G=\frac{\alpha}{3g}\frac{Vh}{U}+ \sigma q\bigg(\frac{U}{gh}- \frac{\psi}{\bar{u}\varphi}\bigg)\,.
   \end{array} \]
The variables describing the lower dispersive layer $U$, $h$ and `internal' variables $V$ and $\zeta$ are always continuous in flow. However, the variables describing the upper turbulent layer may be discontinuous when the supercritical--subcritical transition occurs. Stationary solutions of system~(\ref{eq:model-GLCh}) are obtained and studied in \cite{G_L_Ch_2016}. Below we show that such solutions can be constructed in the framework of model~(\ref{eq:hyp-appr}). Here we restrict our consideration to smooth solutions. 

Further we are looking for stationary solutions of equations~(\ref{eq:st-norm}) having supercritical constant potential flow as $x\to-\infty$:
\begin{equation} \label{eq:const-flow}  
 \begin{array}{l}\displaystyle 
  (h, U) \to (H_0, U_0), \quad (\eta, \bar{u}, q) \to (0, U_0, 0), \\[4mm]\displaystyle  
  (\zeta, V) \to (H_0, 0), \quad U_0>0, \quad F=\frac{U_0}{\sqrt{gH_0}}>1. 
 \end{array} 
\end{equation}   
Here $H_0$ and $U_0$ are the given flow depth and velocity at infinity and $F$ is the Froude number. To construct such a solution, it is necessary first to understand the asymptotic behaviour of the supercritical solution at negative infinity. Linearising equations~(\ref{eq:st-hyp-appr}), we obtain
\begin{equation} \label{eq:st-lin}   
 \begin{array}{l}\displaystyle 
  U_0\tilde{h}'+H_0\tilde{U}'=-\sigma \tilde{q}, \quad 
  U_0\tilde{U}'+\Big(g+\frac{\alpha}{3}H_0\Big)\tilde{h}'+g\tilde{\eta}'+ \frac{\alpha}{3}H_0\tilde{\zeta}'=0, \\[4mm]\displaystyle
  U_0\tilde{\eta}'=\sigma\tilde{q}, \quad U_0\tilde{\bar{u}}'+g\tilde{h}'+g\tilde{\eta}'= \frac{\sigma\tilde{q}}{\tilde{\eta}}(\tilde{U}-\tilde{\bar{u}}), \\[4mm]\displaystyle 
  U_0\tilde{q}'= \frac{\sigma}{2\tilde{\eta}}\big((\tilde{U}-\tilde{\bar{u}})^2-(1+\kappa)\tilde{q}^2\big), \quad 
  U_0\tilde{\zeta}'=\tilde{V}, \quad U_0\tilde{V}'=\alpha(\tilde{h}-\tilde{\eta}). 
 \end{array} 
\end{equation}   
Here and below, the `tilde' symbol denotes small perturbations of the corresponding variables:
\[ \begin{array}{l}\displaystyle  
    h=H_0+\tilde{h}, \quad U=U_0+\tilde{U}, \quad \eta=\tilde{\eta}, \\[2mm]\displaystyle  \bar{u}=U_0+\tilde{\bar{u}}, \quad     
    q=\tilde{q}, \quad \zeta=H_0+\tilde{\zeta}, \quad V=\tilde{V}. 
   \end{array} \]
Following \cite{L_T_2000}, we are looking for the solutions of Eqs.~(\ref{eq:st-lin}) that vanish at negative infinity in the form
\[ (\tilde{h}, \tilde{U}, \tilde{\eta}, \tilde{\bar{u}}, \tilde{q}, \tilde{\zeta}, \tilde{V}) = 
   (\hat{h}, \hat{U}, \hat{\eta}, \hat{\bar{u}}, \hat{q}, \hat{\zeta}, \hat{V}) 
   \exp\Big(\frac{\nu x}{H_0}\Big). \]
Substituting this representation of the solution into Eqs.~(\ref{eq:st-lin}) and expressing the unknown amplitudes (denoted with the ‘hat’ symbol) through $\hat{U}$ and $\nu$, we get
\begin{equation} \label{eq:st-hat}  
 \begin{array}{l}\displaystyle 
  \hat{\bar{u}}=\bigg(1+\frac{1}{F^2}\bigg)\frac{\hat{U}}{2}, 
  \quad \hat{q}=-\frac{\hat{U}}{2\sqrt{3+\kappa}}\bigg(1-\frac{1}{F^2}\bigg), \quad 
  \hat{\eta}=\frac{\sigma H_0}{\nu U_0}\hat{q}, \\[4mm]\displaystyle
  \hat{h}=-\hat{\eta}-\frac{H_0}{U_0}\hat{U}, \quad \hat{\zeta}=\frac{\alpha_1\hat{h}}{\alpha_1+\nu^2 F^2}, 
  \quad \hat{V}=\frac{\nu U_0}{H_0}\hat{\zeta}\,.
 \end{array} 
\end{equation}
Here $\alpha_1=\alpha H_0/g$ and $\hat{U}\leq 0$ is a given amplitude of the small perturbation of velocity. A non-trivial ($\hat{U}\neq 0$) solution of (\ref{eq:st-lin}) in the form~(\ref{eq:st-hat}) exists if the parameter $\nu$ satisfies the equation 
\[ F^2\nu^2+\alpha_1+\frac{((F^2-1)l_2\nu+\alpha_1)l_1\alpha_1}{F^2-1-l_1\alpha_1}=0 \quad\quad 
   \bigg(l_1=\frac{1}{3}, \quad l_2=\frac{\sigma}{2\sqrt{3+\kappa}}\bigg)\,. \]
If the parameter $\alpha_1$ is such that 
\begin{equation} \label{eq:st-alpha} 
  \alpha_1>\frac{4(F^2-1)F^2}{((F^2-1)l_1 l_2^2+4F^2)l_1}=\alpha_1^* 
\end{equation}
(or $\alpha>\alpha_1^*g/H_0$) then for a given Froude number $F>1$ there is a single positive root $\nu$ of the previous quadratic equation
\begin{equation} \label{eq:st-lam} 
  \nu=\frac{(F^2-1)l_1 l_2\alpha_1-\sqrt{(F^2-1)\alpha_1 W}}{2(F^2-1-l_1\alpha_1)F^2}, 
\end{equation}
where $W=(l_1^2 l_2^2\alpha_1+4 l_1\alpha_1+4)F^2-4F^4-l_1^2 l_2^2\alpha_1$. We use asymptotic expressions (\ref{eq:st-hat}), (\ref{eq:st-lam}) when the conditions are imposed at $x=x_0$ in the numerical treatment of  stationary system~(\ref{eq:st-norm}). It should be noted that the amplitudes $\tilde{\bar{u}}$, $\tilde{q}$, $\tilde{\eta}$ and $\tilde{h}$ coincide with the obtained in \cite{G_L_Ch_2016} for asymptotic behaviour of stationary solutions in the framework of Eqs.~(\ref{eq:model-GLCh}), but $\nu$ is different. To solve ODE~(\ref{eq:st-norm}) numerically, we use the standard $ode45$ procedure of the MATLAB package.

\begin{figure}[t]
\begin{center}
\resizebox{1\textwidth}{!}{\includegraphics{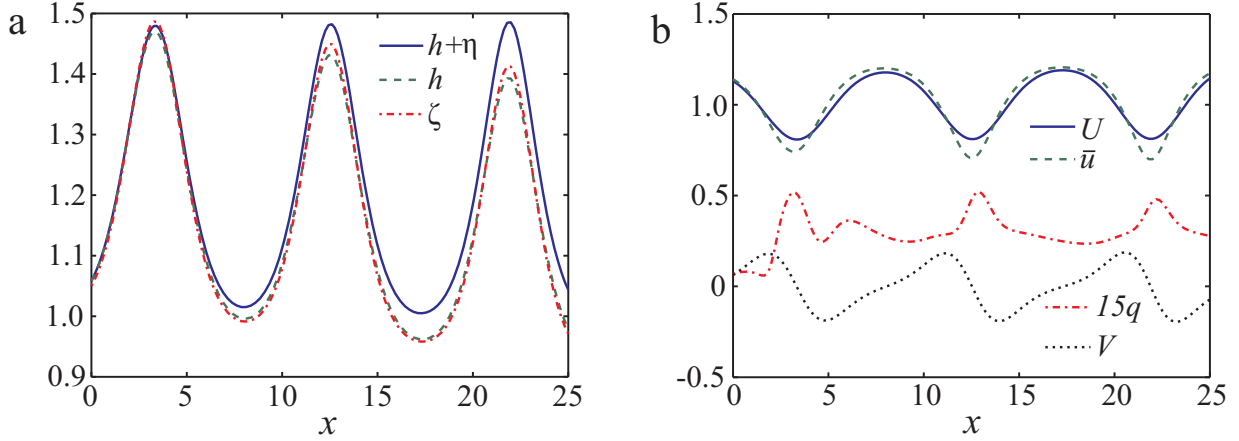}}\\[0pt]
{\caption{Stationary solution for $F=1.2$: (a) --- free surface $h+\eta$, depth of the lower layer $h$ and its instantaneous depth $\zeta$; (b) --- velocity of the layers $U$ and $\bar{u}$, shear velocity $q$ of the upper layer (with factor 15) and instantaneous variable $V$. The parameters are as follows: $g=1$, $\sigma=0.15$, $\kappa=3$ and $\alpha=10$.} \label{fig:fig_stationary}} 
\end{center}
\end{figure}

An example of a stationary solution in the form of an undular bore is shown in Fig.~\ref{fig:fig_stationary}. We choose the parameters $H_0=1$ and $U_0=1.2$ for fluid flow at the negative infinity and solve numerically ODE~(\ref{eq:st-norm}) with conditions at $x=0$ perturbed according to formulas~(\ref{eq:st-hat}) and (\ref{eq:st-lam}) (we take here $\hat{U}=0.07$). The empirical constants, parameter $\alpha$ and non-dimensional gravity acceleration are as follows: $\sigma=0.15$, $\kappa=3$, $\alpha=10$ and $g=1$. In view of restriction (\ref{eq:st-alpha}) for the Froude number $F=1.2$ we have to choose $\alpha>1.32$. In this case the upper turbulent layer develops quite slowly (Fig.~\ref{fig:fig_stationary}(a)) since the flow is supercritical everywhere (the determinants $\varphi$ and $\Delta$ in (\ref{eq:st-norm}) are positive). We also note that the instantaneous depth $\zeta$ of the lower dispersive layer almost coincides with $h$. The larger $\alpha$, the better $\zeta$ approximates the depth of the lower potential layer $h$. The velocities in the layers $U$, $\bar{u}$ and the shear velocity $q$ of the turbulent layer are shown in Fig.~\ref{fig:fig_stationary}(b). Note that in the vicinity of the wave crests, the variable $q$ reaches its maximum value, while the velocities $U$ and $\bar{u}$ are minimal. The smaller the Froude number $F>1$, the larger is the domain where the solution is supercritical. For considered example with $F=1.2$ the supercritical–subcritical transition happens at $x\approx 30.5$,  i.e. the determinant $\Delta$ in (\ref{eq:st-norm}) changes sign from positive to negative. This leads to a jump of the variables of the upper turbulent layer. 

System (\ref{eq:st-hyp-appr}) can be used to describe the structure of turbulent bore if the dispersion effects are negligible ($\alpha=0$). Let the flow depth and velocity be the following $h=H_0$, $u=U_0$ and $\eta=0$ for $x<x_0$. At the point $x=x_0$ the upper turbulent layer is formed ($\eta>0$ for $x>x_0$). The values of the functions at $x=x_0$ are denoted by the subscript ``0''. We choose here  $\zeta_0=H_0$, $V_0=0$ and $\alpha=0$. From the two last equations in (\ref{eq:st-hyp-appr}) we get $\zeta=\zeta_0$, $V=0$. Let us assume that as $x\to x_ 0$ (or $\eta\to 0$) there exist finite limits of the functions $\bar{u}\to\bar{u}_0$ and $q\to q_0$. Assuming that the unknown functions and their derivatives are bounded, from the third, fourth and fifth equations of system~(\ref{eq:st-hyp-appr}) we have 
\[ \eta'\to \frac{\sigma q_0}{\bar{u}_0}, \quad \eta'\to \frac{\sigma}{q_0}(U_0-\bar{u}_0),    \quad (U_0-\bar{u}_0)^2=(1+\kappa)q_0^2.  \]
After eliminating the limiting value $\eta'$ from these relations, we obtain
\begin{equation} \label{eq:st-cond}  
  \bar{u}_0=\frac{U_0}{2+\kappa}, \quad q_0=\frac{\sqrt{1+\kappa}}{2+\kappa}U_0.
\end{equation}
We show below that Eqs.~(\ref{eq:st-norm}) with $\alpha=0$ and conditions~(\ref{eq:st-cond}) describe the stationary turbulent bore when the hydrostatic approximation can be applied \cite{Liapidevskii_Chesnokov_2014}. 

\section{Numerical results}

In this section, we present the results of numerical simulation of the formation and evolution of turbulent bores. 
First of all, we show that terms of order $O(\sigma\varepsilon^2)$ practically do not affect the numerical results in framework of model~(\ref{eq:model-GLCh}). Then, we clearly demonstrate on an example of Favre waves that as the $\alpha$ parameter increases, the solutions of the hyperbolic system~(\ref{eq:hyp-appr}) converge to the solutions of the corresponding dispersive equations. We also present the results of modelling of undular and monotone bores formed as a result of the flow past an obstacle. In the end of this section we consider the shoaling and breaking of a solitary wave. 

To solve differential balance laws~(\ref{eq:hyp-appr}) numerically, we implement here the Nessyahu--Tadmor second-order central scheme~\cite{N_T_1990}. The same method is used to solve dispersive equations~(\ref{eq:model-GLCh}) for the unknown functions $\mathbf{u}=(h, \eta, K, \bar{u}\eta, q)$, but at each time step we apply the Thomas algorithm (a simplified form of Gaussian elimination) to find the velocity $U$ in the lower layer from second-order equation~(\ref{eq:K-def}). 

\subsection{Favre waves}

With accuracy up to the terms of order $O(\sigma\varepsilon^2)$, system (\ref{eq:model-GLCh}) for the evolutionary variables $\mathbf{u}=(h, \eta, K, \bar{u}\eta, q)$ can be written as
\begin{equation}\label{eq:model-GLCh-mod}
 \begin{array}{l}\displaystyle
  h_t+ (Uh)_x=-\sigma q, \quad \eta_t+(\bar{u}\eta)_x=\sigma q, \\[3mm]\displaystyle 
  K_t +\Big( KU +g(h+\eta) -\frac{U^2}{2} -\frac{\varepsilon^2U_x^2 h^2}{2} \Big)_x = -gb_x, \\[3mm]\displaystyle 
  (\bar{u}\eta)_t+ \Big((\bar{u}^2+q^2)\eta+\frac{g\eta^2}{2}\Big)_x=\sigma qU- g\eta h_x -g\eta b_x, \\[3mm]\displaystyle 
  q_t+(\bar{u}q)_x= \frac{\sigma}{2\eta}\big((U-\bar{u})^2-(1+\kappa)q^2\big) 
 \end{array}
\end{equation}
with equation~(\ref{eq:K-def}) for the variable $U$ to be solved at each time step. Let us show that there is almost no difference in calculations on the basis of equations~(\ref{eq:model-GLCh-mod}), (\ref{eq:K-def}) and their original version used in \cite{G_L_Ch_2016}, where equation~(\ref{eq:K}) stands for the third equation in system~(\ref{eq:model-GLCh-mod}). To do this, we repeat the test concerning Favre waves considered in \cite{G_L_Ch_2016}.

\begin{figure}[t]
\begin{center}
\resizebox{1\textwidth}{!}{\includegraphics{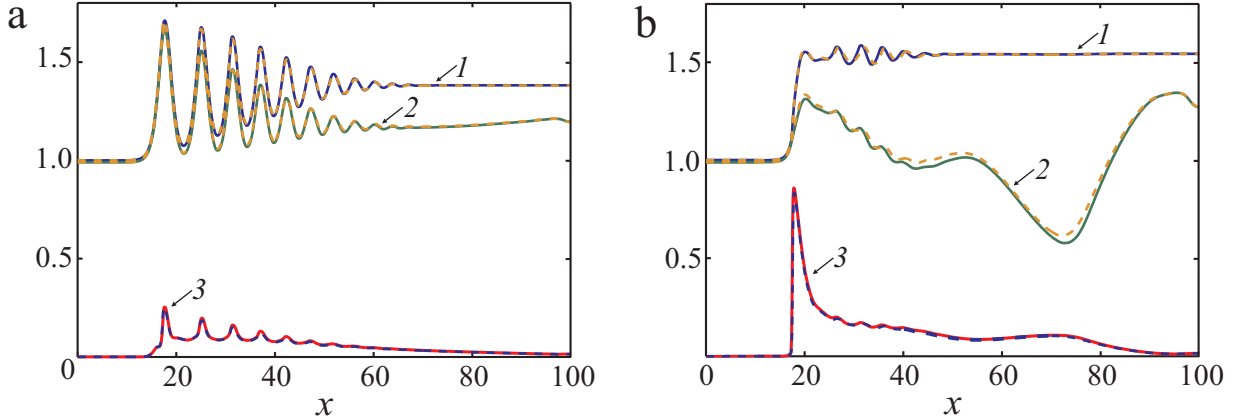}}\\[0pt]
{\caption{Favre waves: {\it 1} --- free surface $h+\eta$, {\it 2} --- thickness of the lower layer $h$, and {\it 3} --- variable $q^*=q/U_0$ at $t=90$ for the wave Froude numbers $F_w=1.28$ (a) and $F_w=1.40$ (b). Solid curves --- model~(\ref{eq:model-GLCh-mod}), dashed --- equations~(\ref{eq:model-GLCh}).} \label{fig:fig_Favre_1}} 
\end{center}
\end{figure}

We perform calculations in dimensionless variables in the domain $x\in [0,100]$ on a uniform grid with $N=1000$ nodes. To start calculations we specify unknown variables $\mathbf{u}$ in the node points $x=x_j$ at $t=0$ as follows $\eta=0.01$, $h+\eta=1$, $\bar{u}=K=U_0$, and $q=0$. The right boundary $x=100$ is an impermeable wall. To satisfy the impermeability condition ($\bar{u}=U=0$) we set $\bar{u}_{N+1}=-\bar{u}_{N-1}$ and $K_{N+1}=-K_{N-1}$. At $x=0$ the initial data are used as boundary condition. We choose here $g=1$, $\varepsilon=1$, $\sigma=0.15$ and $\kappa=3$. 

Instead of the upstream velocity $U_0$, it is more convenient to prescribe the wave Froude number $F_w$:
\[ F_w=\frac{U_0-D}{\sqrt{gH_0}}\,, \]
where $D$ denotes the propagation speed of the reflected wave. A simple relation between $U_0$ and $\bar{F}$ can be obtained in the form (see \cite{G_L_Ch_2016} for details)
\[ U_0=\sqrt{gH_0}\bigg(F_w-\frac{1+\sqrt{1+8F_w^2}}{4F_w}\bigg)\,. \] 
In supercritical regime ($F_w>1$), undulations start developing at the bore front. When the wave Froude number is approximately  between 1.3 and 1.4, the transition from the undular bore to the breaking bore occurs. This problem was studied in experimental works \cite{Favre, Treske_1994} and numerically \cite{Tissier_2012, G_L_Ch_2016}. 

The results of calculations for the wave Froude numbers $F_w=1.28$ and $F_w=1.40$ are shown in Fig.~\ref{fig:fig_Favre_1}. We clearly see that terms of order $O(\sigma\varepsilon^2)$ in (\ref{eq:model-GLCh}) almost do not affect on the numerical result, especially for an undular bore. A similar coincidence is observed for the other tests considered \cite{G_L_Ch_2016} including the shoaling and breaking of a solitary wave propagating in a long channel of mild slope. Thus, for carrying out calculations, the two-layer model~(\ref{eq:model-GLCh}) can be used in simplified form~(\ref{eq:model-GLCh-mod}), (\ref{eq:K-def}).

\begin{figure}[t]
\begin{center}
\resizebox{1\textwidth}{!}{\includegraphics{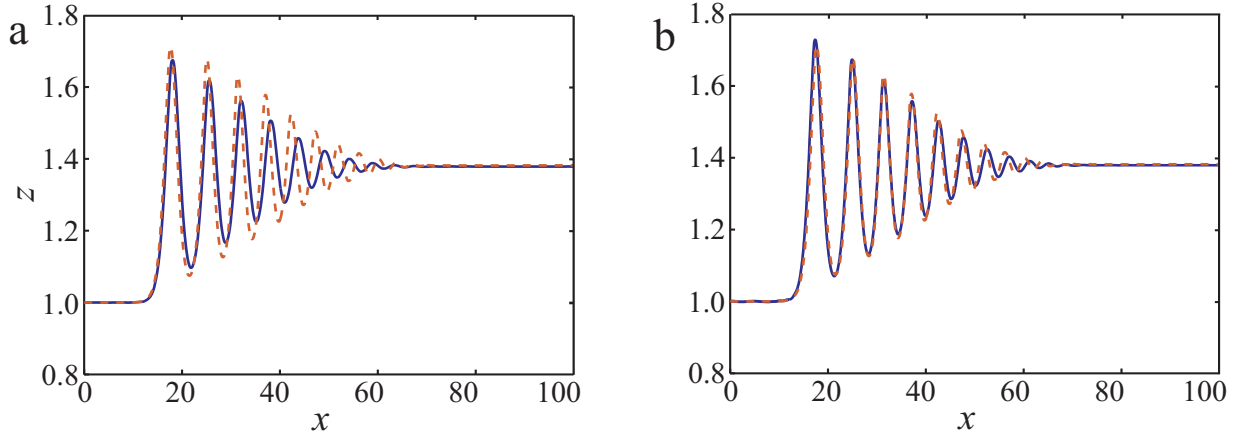}}\\[0pt]
{\caption{Favre waves: free surface $h+\eta$ at $t=90$ for $F_w=1.28$. Solids --- solution of Eqs.~(\ref{eq:hyp-appr}) obtained for $N=1000$, $\alpha=10$ (a) and $N=1700$, $\alpha=20$ (b). Dashed --- solution of Eqs.~(\ref{eq:model-GLCh-mod}) obtained with resolution $N=1000$. The other parameters correspond to Fig.~\ref{fig:fig_Favre_1}.} \label{fig:fig_Favre_2}} 
\end{center}
\end{figure}

Let us show that the solutions of the proposed hyperbolic model~(\ref{eq:hyp-appr}) approximate the solutions of dispersive equations~(\ref{eq:model-GLCh-mod}), (\ref{eq:K-def}). Fig.~\ref{fig:fig_Favre_2}(a) presents free surface $z=h+\eta$ for the previous test for $F_w=1.28$ obtained on the basis of system~(\ref{eq:hyp-appr}) with $\alpha=10$ (solid curve) and in the framework of Eqs.~(\ref{eq:model-GLCh-mod}), (\ref{eq:K-def}) (dashed curve). For the hyperbolic equations~(\ref{eq:hyp-appr}) at $t=0$ we add conditions for the `instantaneous' variables as follows $\zeta=h$ and $V=0$. There is a good match for the leading wave, while for the secondary waves some difference is observed. An almost complete coincidence of the solutions of the dispersion model and its hyperbolic approximation can be achieved by increasing the parameter $\alpha$ (see Fig.~\ref{fig:fig_Favre_2}(b)). However, for this, the spatial resolution should be improved the calculation by using the hyperbolic model~(\ref{eq:hyp-appr}). This is due to the fact that as $\alpha $ increases, the velocity of the characteristics (in absolute value) also increases and, consequently, the time step becomes smaller. As a result, it leads to a stronger effect of the numerical viscosity. 

For other Froude numbers and, in particular, in the transition to monotonous bore, solutions of hyperbolic equations~(\ref{eq:hyp-appr}) also approximate solutions of the dispersion model.

It is important to note that the verification of the dispersive model (\ref{eq:model-GLCh}) performed in \cite{G_L_Ch_2016} is based on a comparison of the maximum and minimum amplitude of the first wave with the experimental data \cite{Favre} and \cite{Treske_1994}. As we can see from Fig.~\ref{fig:fig_Favre_2}, the results of calculations for the first wave by using model (\ref{eq:hyp-appr}) with $\alpha=10$ and $\alpha=20$ almost coincide, so there is no need to choose large values of $\alpha$. In addition, a finer mesh is necessary for large values of the relaxation parameter $\alpha$. 

\subsubsection{Dependence of the solution on the empirical parameters}

Both models (\ref{eq:model-GLCh-mod}) and (\ref{eq:hyp-appr}) include the empirical parameters $\sigma$ and $\kappa$, which may differ slightly from the fixed above values. In particular, the following range of these parameters were used in \cite{L_T_2000, Liapidevskii_Chesnokov_2014, G_L_Ch_2016} for layered hydrostatic and dispersive models: $\sigma\in [0.15, 0.20]$ and $\kappa\in [2, 6]$. A change in the empirical parameters in the specified range does not lead to a significant change in the solution. Let us consider this in more detail.  

\begin{figure}[t]
\begin{center}
\resizebox{1\textwidth}{!}{\includegraphics{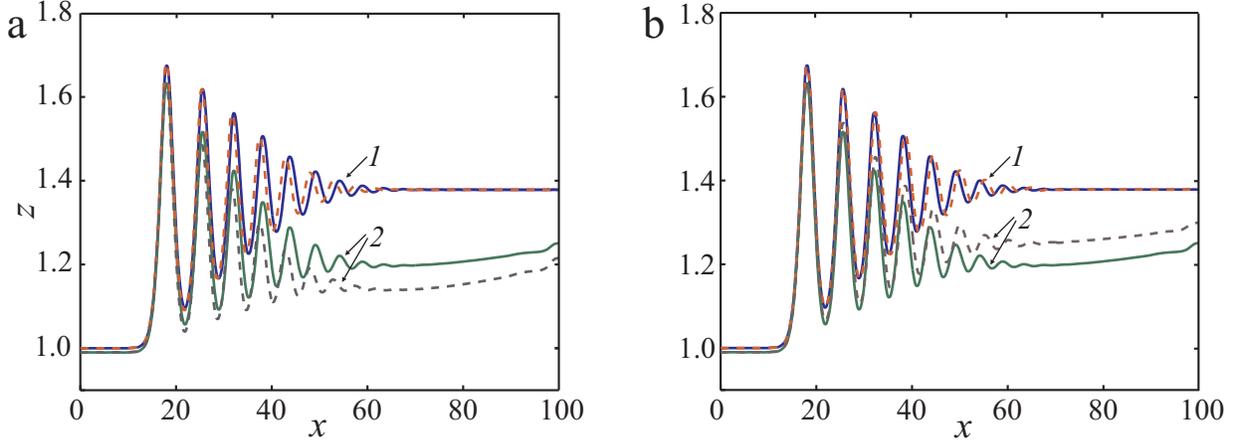}}\\[0pt]
{\caption{Favre waves: {\it 1} --- free surface $h+\eta$ and {\it 2} --- thickness of the lower layer $h$ at $t=90$ for the wave Froude numbers $F_w=1.28$ obtained by model (\ref{eq:hyp-appr}) with $\alpha=10$. Solid curves --- $\sigma=0.15$ and $\kappa=3$; dashed --- $\sigma=0.20$ and $\kappa=3$ ($a$); $\sigma=0.15$ and $\kappa=6$ ($b$).} \label{fig:fig_Favre_3}} 
\end{center}
\end{figure}

Fig.~\ref{fig:fig_Favre_3} shows the results of the Favre waves calculation ($F_w=1.28$) obtained with the help of hyperbolic model (\ref{eq:hyp-appr}) for different empirical parameters $\sigma$ and $\kappa$. As can be seen from Fig.~\ref{fig:fig_Favre_3}(a), an increase in the parameter $\sigma$ leads to a decrease in the thickness $h$ of the lower layer, as well as to an insignificant decrease in the wavelength (dashed curves). At the same time, the fluid depth $h+\eta $, in contrast to the `artificial' internal interface, almost coincides with the case $\sigma=0.15$ and $\kappa=3$ (solid curves), especially for the leading wave. In the case of an increase in the parameter $\kappa$ (see Fig.~\ref{fig:fig_Favre_3}(b)), the thickness of the lower layer increases and the wavelength is slightly longer. However, in the vicinity of the leading wave the difference in the results is minimal. Thus, we can use fixed parameters $\sigma=0.15$ and $\kappa=3$. Moreover, there is no need to change these parameters for other tests.

\subsubsection{A comparison of computational time for the hyperbolic and dispersive models}

Let us find out the dependence of the computational time for model (\ref{eq:hyp-appr}) on the value of the relaxation parameter $\alpha$. We choose the same data as at the beginning of this section, in particular, $F_w=1.28$, $N=1000$ and $t_{\max}=90$. It is obviously, the computation time depends on the velocities of the characteristics, since the time step is determined by the Courant condition
\begin{equation}\label{eq:CFL} 
  {\rm CFL}=\frac{\Delta t}{\Delta x} \max_{k,j}|\lambda_k(\bU_j^n)|<\frac{1}{2}\,, 
\end{equation}
where $\Delta x$ is the spatial grid spacing, $\lambda_k$ is the characteristic velocity and $\bU_j^n$ is the numerical solution at $t=t^n$ and $x=x_j$. In what follows, we take ${\rm CFL}=0.475$. Note that $\alpha=0$ corresponds to a hydrostatic approximation. In this case, the calculation time is minimal, since the maximum value of the characteristic velocity increases with the relaxation parameter $\alpha$. 

The calculation time $T_\alpha$ (in seconds) for this test on the computer being used is presented in Table \ref{tab:Test1} for different values of $\alpha$. We also indicate here the number of iterations $M_{\alpha}$ (time steps) needed to reach the final time $t_{\max}$. This parameter is determined only by the used numeric code and does not depend on the CPU power. As we can see from Table~\ref{tab:Test1}, the computation time and the number of iterations increase with the relaxation parameter $\alpha$. However, the increment in computation time (or number of iterations) decreases with the growth of $\alpha$. In particular, $T_6-T_3=2.5382$ ($M_6-M_3=1083$), $T_{12}-T_9=1.4285$ ($M_{12}-M_9=737$). A further increase in the parameter $\alpha$ at the given grid resolution $N$ is impractical, since an increase in the number of steps in time enhances the numerical viscosity. Therefore, it is necessary to improve the spatial resolution, as it was done in the example shown in Fig.~\ref{fig:fig_Favre_2}(b). 

For the numerical solution of the dispersive system (\ref{eq:model-GLCh-mod}), we employ the approach proposed in \cite{LeM_G_H_2010} with the inversion of the elliptic operator (\ref{eq:K-def}) at each time step. Although model (\ref{eq:model-GLCh-mod}) is not a hyperbolic system, we apply condition (\ref{eq:CFL}) to determine the time step $\Delta t$. We take the velocity of the characteristic $\max\limits_k|\lambda_k|$ of the hydrostatic model ($\varepsilon=0$) with factor $a_d>1$ needed to stabilize the numerical algorithm. We note that system (\ref{eq:model-GLCh-mod}) contains a term of the form $\varepsilon^2(h^2U_x)_x$. Therefore, to stabilize the numerical code the velocity of the characteristic should be larger. The choice of the factor $a_d$ depends on the considered problem. Here we take $a_d=3/2$. In this case the computation time is $T=12.1951$ seconds and it takes $M=3840$ iterations to reach the final time $t_{\max}$. 

\begin{table}[t]
\begin{center}
\begin{tabular}{|c|c|c|c|c|c|} \hline
               & $\alpha=0$ & $\alpha=3$ & $\alpha=6$ & $\alpha=9$ & $\alpha=12$ \\ \hline 
$T_\alpha$     & 5.8660     & 8.0554     & 10.5936    & 12.8388    & 14.2673     \\ \hline 
$M_{\alpha}$            & 2560       & 3483       & 4566       & 5446       & 6183        \\ \hline
\end{tabular}
\end{center}
\caption{Computational time $T_{\alpha}$ and number of iterations (time steps) $M_{\alpha}$ of model (\ref{eq:hyp-appr}) for different values of the relaxation parameter $\alpha$.} \label{tab:Test1}
\end{table}

Comparing with the data from Table~\ref{tab:Test1}, we see that the calculation time for the dispersion model (\ref{eq:model-GLCh-mod}) approximately corresponds to the hyperbolic model with $\alpha=9$, despite the greater number of iterations required for system (\ref{eq:hyp-appr}). The computational time spent per iteration for equations (\ref{eq:model-GLCh-mod}) is longer than for system (\ref{eq:hyp-appr}). It is clear, if the spatial resolution improves, the time spent on iteration for the dispersive model increases significantly due to the inversion of the elliptic operator (\ref{eq:K-def}). In this sense the hyperbolic model works faster for all reasonable $\alpha$. In most cases $\alpha \in [5,8]$ provides a fairly good approximation, and the computation time for the hyperbolic model is shorter, despite the greater number of iterations. 

Let us note that $\alpha$ should be multiplied by $g/H_0$ if we use dimensional variables.

\subsection{Flows over a local obstacle}

Let us consider the flows arising from the interaction of the upstream supercritical flow with a local obstacle located in the vicinity of the outlet section of the channel. Here we perform calculations using of the hyperbolic system (\ref{eq:hyp-appr}) in the domain $x\in [0,35]$ and set the constant flow $h_0=\zeta_0=0.99$, $\eta_0=0.01$, $U_0=\bar{u}_0=1.2$, $q_0=V_0=0$ (in this case the Froude number $F=1.2$) as the initial data at $t=0$. These data are used as a boundary condition at $x=0$. On the right boundary we pose the Neumann condition. The relaxation parameter $\alpha$ is equal to 10. As before, we take $g=1$, $\varepsilon=1$, $\sigma=0.15$ and $\kappa=3$.

\begin{figure}[t]
\begin{center}
\resizebox{1\textwidth}{!}{\includegraphics{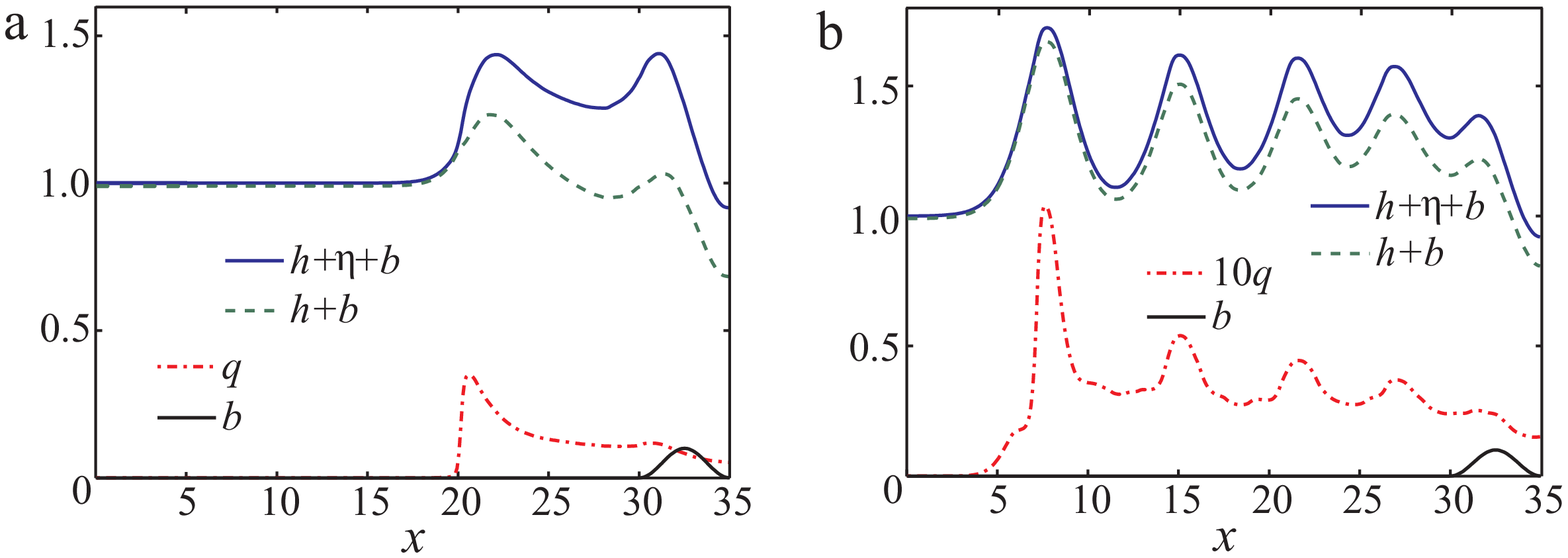}}\\[0pt]
{\caption{Undular bore generated by bottom topography $z=b(t,x)$ and propagated on supercritical flow ($F=1.2$): free surface $z=h+\eta+b$, interface $z=h+b$ and variable $q$ at $t=85$ ($a$) and $t=225$ ($b$).} \label{fig:fig_obstacle_1}} 
\end{center}
\end{figure}

The formation of bore is carried out by controlling the obstacle. At the initial time, the bottom is flat ($b=0$). During time $t_1=25$, the height of the smooth obstacle of width $5$ located on the right edge is increased to $b^*=0.2$, then, over time $t_2=50$ it is reduced to $b_*=0.1$. It leads to the formation of an undular bore shown in Fig.~\ref{fig:fig_obstacle_1}. At the initial stage of the process, the thickness of the upper turbulent layer increases substantially and the shear velocity $q$ sharply increases at the wave front (Fig.~\ref{fig:fig_obstacle_1}(a)). Then a wave train propagating upstream the flow is formed (Fig.~\ref{fig:fig_obstacle_1}(b)). The thickness of the upper turbulent layer becomes insignificant in the vicinity of the leading wave and the flow qualitatively corresponds to the stationary solution (Fig.~\ref{fig:fig_stationary}(a)). Note that the speed of waves propagation depends on the height of the obstacle. It should be greater than a certain limit value for the given number Froude of the upstream flow.

\begin{figure}[t]
\begin{center}
\resizebox{1\textwidth}{!}{\includegraphics{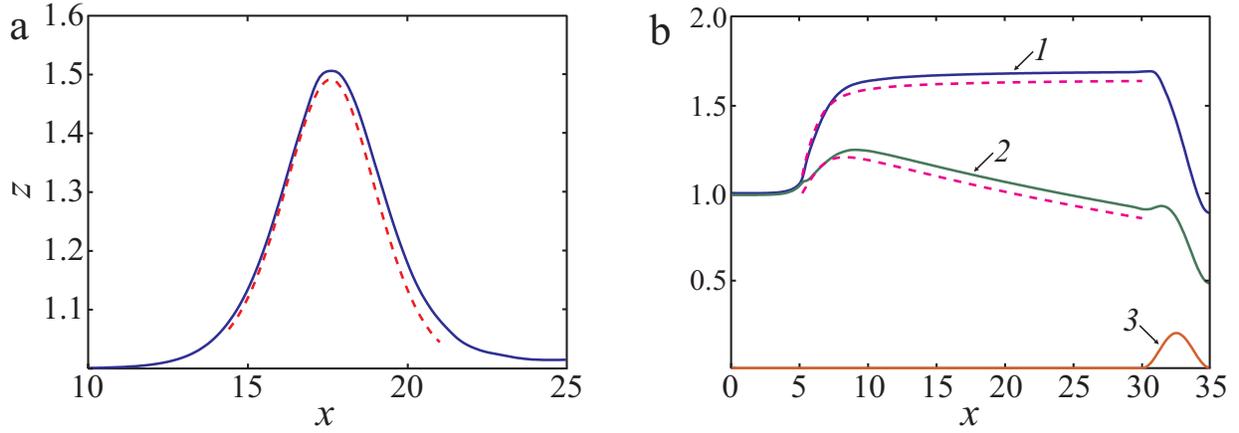}}\\[0pt]
{\caption{Non-stationary (solid curve) and stationary (dashed) solutions of Eqs.~(\ref{eq:hyp-appr}) obtained for $F=1.2$ (a) and $F=1.4$ (b).} \label{fig:fig_obstacle_2}} 
\end{center}
\end{figure}

Let us repeat the previous calculation, with the difference that for $t>t_2=50$ the bottom topography becomes even ($b_*=0$). In this case, a solitary wave slowly propagating upstream is formed. This wave (the free surface $z=h+\eta$) at $t=225$ is shown in Fig.~\ref{fig:fig_obstacle_2}(a) (solid curve). The dashed line corresponds to the solution of stationary Eqs.~(\ref{eq:st-norm}) with the same parameter $\alpha=10$. As can be seen from the figure, the profile of a solitary wave is quite well described by a stationary solution.

With increase in the velocity of upstream flow (and the height of the obstacle), a monotone turbulent bore is formed. For this calculation we choose $t_1=15$, $b^*=0.3$, $t_2=30$, $b_*=0.2$, $U_0=\bar{u}_0=1.4$ and the remaining parameters do not change. The result of calculations on the basis of Eqs.~(\ref{eq:hyp-appr}) at $t=285$ (the free surface $z=h+\eta+b$ and the interface $z=h+\eta$) is shown in Fig.~\ref{fig:fig_obstacle_2}(b) (solid curves {\it 1} and {\it 2}, line {\it 3} --- bottom topography). The solution of stationary equations~(\ref{eq:st-norm}) with $\alpha=0$ (it corresponds to the hydrostatic approximation) and conditions~(\ref{eq:st-cond}) at $x=x_0=5$ is shown by the dashed lines. This solution is in a good agreement with non-stationary calculation. Note that a local area of the subcritical flow is formed above the obstacle (the variable $\Delta$ in (\ref{eq:st-norm}) changes the sign), at the exit from the channel the flow becomes supercritical again. For this reason, the stationary solution is constructed only in the interval before the obstacle. The given example shows that Eqs.~(\ref{eq:hyp-appr}) with a fixed parameter $\alpha$ allow one to describe flows when the effect of dispersion is important (Fig.~\ref{fig:fig_obstacle_2}(a)), as well as flows that can be described in the hydrostatic approximation (Fig.~\ref{fig:fig_obstacle_2}(b)).

\begin{rmk} 
There is an additional advantage of the hyperbolic approximation (\ref{eq:hyp-appr}), which consists in the formulation of boundary conditions. As a rule, when implementing Godunov-type numerical schemes, the formulation of boundary conditions depends on the number of incoming and outgoing characteristics at the boundary of the computational domain. Due to seven equations in system (\ref{eq:hyp-appr}), it is convenient to use central schemes \cite{N_T_1990}, which do not require exact or approximate solution of the Riemann problem. In this case we have to specify the boundary conditions for all unknown functions. Let the flow at the exit from the computational domain be supercritical and the disturbances propagating upstream do not reach the left boundary of the region, as in the example with the flow over an obstacle. Then the Neumann conditions at the right boundary do not affect the solution inside of the computational domain. It is not valid for the dispersive model since at each time step a non-local operation of inversion of the elliptic operator is required. As a result, the boundary conditions affect the solution. 
\end{rmk}

\subsection{The breaking of a solitary wave}

The evolution of breaking solitary wave on a mild sloping beach in the framework of the two-layer dispersive model (\ref{eq:model-GLCh}) was considered in \cite{G_L_Ch_2016}, where an excellent agreement with experimental data \cite{Hsiao} was shown. Therefore, to verify the proposed hyperbolic model (\ref{eq:hyp-appr}) in this test, it suffices to compare the numerical results with calculations using of dispersive equations (\ref{eq:model-GLCh}).

\begin{figure}[t]
\begin{center}
\resizebox{.7\textwidth}{!}{\includegraphics{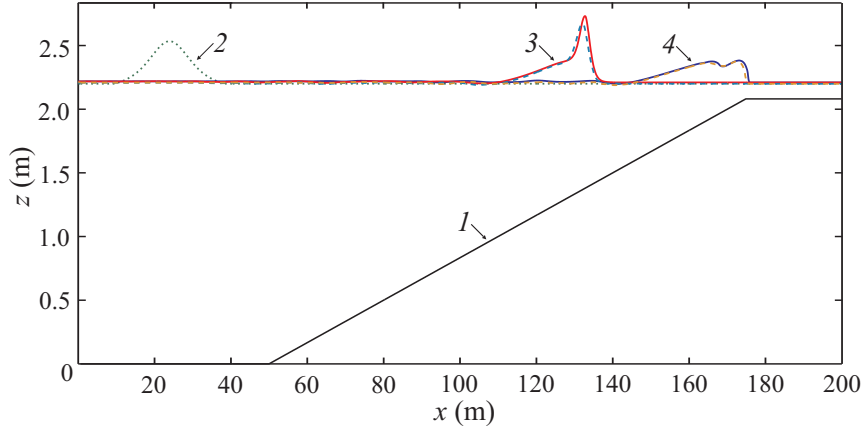}}\\[0pt]
{\caption{The evolution of solitary wave on a mild sloping beach: {\it 1} --- bottom topography; {\it 2, 3}, and {\it 4} --- free surface at $t_*=0, \, 50$, and $75$. Here the dimensionless time $t_*$ is given by $t_*=t\sqrt{g/H_0}$. Solid curves --- dispersive model (\ref{eq:model-GLCh}), dashed curves --- hyperbolic model (\ref{eq:hyp-appr}) with $\alpha=11g/H_0$.} \label{fig:fig_beach_1}} 
\end{center}
\end{figure}

Here we perform the calculations in dimensional variables ($g=9.8$\,m/s$^2$) in the domain $x\in [0,200]$ using of $N=1500$ nodes for space resolution. As in the previous tests we take $\sigma=0.15$ and $\kappa=3$. The bottom topography has a slope of 1/60 (curve {\it 1} in Fig.~\ref{fig:fig_beach_1}). To avoid modelling of the wave propagation over dry bottom (run-up), we introduce a shelf zone near the right boundary. Initial data represent a solitary wave having amplitude $a_0=0.3344$\,m propagating with the velocity $C_0=\sqrt{g(a_0+H_0)}$, where $H_0=2.2$\,m is the undisturbed water depth. The initial thickness of the upper turbulent layer $\eta_0$ is equal to $0.01$\,m. On the left and right walls we assume reflecting boundary conditions and take $\alpha=11g/H_0$ for system (\ref{eq:hyp-appr}). Fig.~\ref{fig:fig_beach_1} shows the results of calculations of the free surface at $t_*=t\sqrt{g/H_0}=50$ and $t_*=75$ with respect to models (\ref{eq:model-GLCh}) and (\ref{eq:hyp-appr}) (solid and dashed curves, correspondingly). 

\begin{figure}[t]
\begin{center}
\resizebox{1\textwidth}{!}{\includegraphics{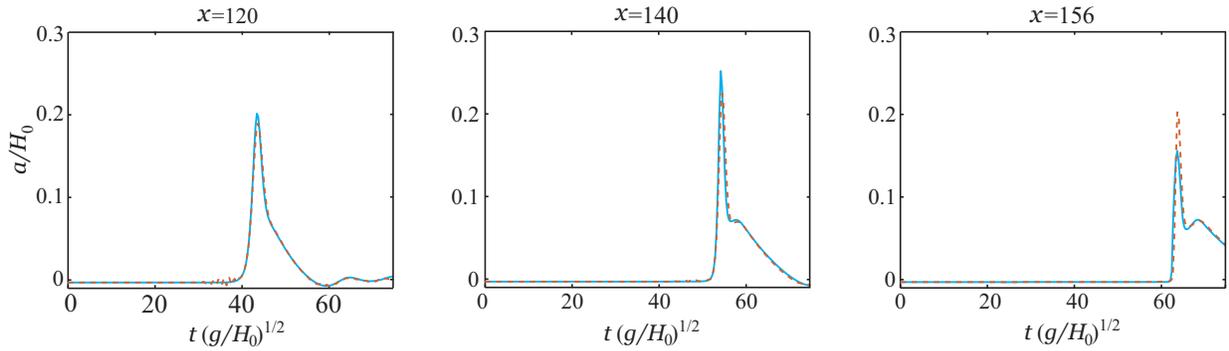}}\\[0pt]
{\caption{Time history of free-surface evolution: solid curves --- model (\ref{eq:model-GLCh}), dashed curves --- model (\ref{eq:hyp-appr}).} \label{fig:fig_beach_2}} 
\end{center}
\end{figure}

The temporal evolution of the free-surface displacements at different positions along the channel is shown in Fig.~\ref{fig:fig_beach_2}. Here we also compare the calculation results for models~(\ref{eq:model-GLCh}) and (\ref{eq:hyp-appr}). The original dispersive model was verified in \cite{G_L_Ch_2016} by experimental data \cite{Hsiao} for the time history of the free surface. Therefore, it can be said that hyperbolic model~(\ref{eq:hyp-appr}) also gives an excellent agreement with the experiment \cite{Hsiao} for the problem of the breaking of a solitary wave. 

It should be noted that it is impossible to model the run-up and run-down using governing equations (\ref{eq:model-GLCh}) or (\ref{eq:hyp-appr}) because a singularity occurs after the upper turbulent layer reaches the bottom. Modifying the systems to describe such process is an open problem.

\section{Conclusion} 

We derived hyperbolic equations (\ref{eq:hyp-appr}) approximating the two-layer long-wave model (\ref{eq:model-GLCh}) proposed in \cite{G_L_Ch_2016} and taking into account dispersion and vortex effects. The construction of the hyperbolic model is based on the use of additional `instantaneous' variables describing the flow of a dispersive fluid layer~\cite{L_T_2000, Liapidevskii_Gavrilova_2008}. The advantage of the hyperbolic model is simpler numerical implementation and formulation of boundary conditions. In contrast to the solution of dispersive equations of the Green--Naghdi type, in this case there is no need for a time-consuming operation for inverting an elliptic operator at each time step. 

The velocities of characteristics of the proposed equations (\ref{eq:hyp-appr}) are determined and hyperbolicity of this system (for sufficient large relaxation parameter) is established. Linear analysis of the derived and original models is performed and the dispersive relations are obtained. A comparison of these relations allows us to observe the effect of the relaxation parameter on the accuracy of the hyperbolic approximation. Stationary solutions of the hyperbolic system (\ref{eq:hyp-appr}) are studied. In particular, we find the asymptotic behaviour of the stationary solution at infinity and construct a solution describing an undular bore with a turbulent sub-surface layer (Fig.~\ref{fig:fig_stationary}). Further, we apply the hyperbolic system to model Favre waves, non-hydrostatic flows over a local obstacle and the evolution of breaking solitary wave on a sloping beach. 

Firstly, we establish that terms of order $O(\sigma\varepsilon^2)$ in model (\ref{eq:model-GLCh}) have practically no effect on the results of calculations for the Favre waves. Therefore, model (\ref{eq:model-GLCh}) can be used in a simpler form (\ref{eq:model-GLCh-mod}) (see Fig.~\ref{fig:fig_Favre_1}). Then we clearly show that solutions of hyperbolic equations (\ref{eq:hyp-appr}) approximate solutions of the dispersive model (\ref{eq:model-GLCh-mod}) (Fig.~\ref{fig:fig_Favre_2}). It is found that for a good approximation of the leading wave it is not necessary to choose large values of the relaxation parameter. We show that variation of the empirical parameters of the model has little effect on the solution of the equations of motion (Fig.~\ref{fig:fig_Favre_3}). Moreover, these parameters are not required to change for other tests. We also compare the time consuming for the dispersive and hyperbolic models. The formation and evolution of a turbulent bore in the interaction of a constant upstream flow with a local obstacle are considered. It is established that at the initial stage of wave bore formation a near-surface turbulent layer intensively develops. The thickness of the upper turbulent layer becomes smaller at least for the leading wave (Fig.~\ref{fig:fig_obstacle_1}). By controlling the obstacle in the vicinity of the output section of the channel, various wave configurations can be realized. In particular, it is possible to obtain a solitary wave or monotone bore. Such examples are presented in Fig.~\ref{fig:fig_obstacle_2}. It is shown that the structure of these waves can be described by stationary equations. We also present the results of the numerical modelling of the shoaling and breaking of a solitary wave (Fig. \ref{fig:fig_beach_2}) which verified by experimental data. 

The used method for constructing the hyperbolic approximation of the dispersive system can be applied to other models, in particular, to multilayer dispersive equations of stratified flows \cite{G_L_Ch_2019, Liapidevskii_2017}. The construction of such hyperbolic approximations of dispersive equations seems especially promising for the numerical simulation of spatial flows. 

\section*{Acknowledgments}

This work was partially supported by the Russian Foundation for Basic Research (project 19-01-00498). The authors thank V.Yu. Liapidevskii, S.L. Gavrilyuk and I.V. Stepanova for fruitful discussions. They also address special thanks to the reviewers for their helpful comments and suggestions.
% and Interdisciplinary Program of SB RAS (project II.1.2)


\begin{thebibliography}{99}

\bibitem{Besse_2017}
C. Besse, P. Noble, D. Sanchez, Discrete transparent boundary conditions for the mixed KDV--BBM equation, J. Comput. Phys. 345 (2017) 484--509.

\bibitem{Bonneton_2011}
P. Bonneton, F. Chazel, D. Lannes, F. Marche, M. Tissier, A splitting approach for the fully nonlinear and weakly dispersive Green–Naghdi model, J. Comput. Phys. 230 (2011) 1479--1498.

\bibitem{Bradshaw_1967}
P. Bradshaw, D.H. Ferriss, N.P. Atwell, Calculation of boundary -- layer development using the turbulent energy equation, J. Fluid Mech. 28 (1967) 593--616.

\bibitem{Bristeau_2015}
M.-O. Bristeau, A. Mangeney, J. Sainte-Marie, N. Seguin, An energy-consistent depth-averaged Euler system: Derivation and properties, Discrete Contin. Dyn. Syst. Ser. B. 20 (2015) 961--988.

\bibitem{Chesn_2017}
A.A. Chesnokov, G.A. El, S.L. Gavrilyuk, M.V. Pavlov, Stability of shear shallow water flows with free surface,
SIAM J. Appl. Math. 77 (2017) 1068--1087.

\bibitem{Escalante_2019}
C. Escalante, E.D. Fern\'andez-Nieto, T. Morales de Luna, M.J. Castro, An efficient two-layer non-hydrostatic approach for dispersive water waves, J. Sci. Comput. 79 (2019) 273--320.

\bibitem{Favre} 
H. Favre, Ondes de translation dans les canaux d\'{e}couverts, Dunod, Paris, 1935.

\bibitem{Favrie_Gavr_2017}
N. Favrie, S. Gavrilyuk, A rapid numerical method for solving Serre--Green--Naghdi equations describing long free surface gravity waves, Nonlinearity 30 (2017) 2718--2736.

\bibitem{Fernandez-Nieto_2018}
E.D. Fern\'andez-Nieto, M. Parisot, Y. Penel, J. Sainte-Marie, A hierarchy of dispersive layer-averaged approximations of Euler equations for free surface flows, Commun. Math. Sci. 16 (2018) 1169--1202.

\bibitem{G_L_Ch_2016}
S.L. Gavrilyuk, V.Yu. Liapidevskii, A.A. Chesnokov, Spilling breakers in shallow water: applications to Favre waves and to the shoaling and breaking of solitary waves, J. Fluid Mech. 808 (2016) 441--468. 

\bibitem{G_L_Ch_2019}
S.L. Gavrilyuk, V.Yu. Liapidevskii, A.A. Chesnokov, Interaction of a subsurface bubble layer with long internal waves, Europ. J. Mech. B/Fluids 73 (2019) 157--169. 

\bibitem{Gavr_Tesh_2001}
S.L. Gavrilyuk, V.M. Teshukov, Generalized vorticity for bubbly liquid and dispersive shallow water equations, Continuum Mech. Thermodyn. 13 (2001) 365--382.

\bibitem{Givoli_2003}
D. Givoli, B. Neta, High-order non-reflecting boundary conditions for the dispersive shallow water equations, J. Comput. Appl. Math. 158 (2003) 49--60.

\bibitem{Green_Naghdi_1976}
A.E. Green, P.M. Naghdi, A derivation of equations for wave propagation in water of variable depth, J. Fluid Mech. 78 (1976) 237--246.

\bibitem{Grosso_2010}
G. Grosso, M. Antuono, M. Brocchini, Dispersive nonlinear shallow-water equations: some preliminary numerical results, J. Eng. Math. 67 (2010) 71--84.

\bibitem{Hsiao} 
S.-C. Hsiao, T.-W. Hsu, T.-C. Lin, Y.-H. Chang, On the evolution and run-up of breaking solitary waves on a mild sloping beach, Coast. Eng. 55 (2008) 975--988.

\bibitem{Ivanova_2017}
K.A. Ivanova, S.L. Gavrilyuk, B. Nkonga, G.L. Richard, Formation and coarsening of roll-waves in shear shallow water flows down an inclined rectangular channel, Comput. Fluids 159 (2017) 189--203. 

\bibitem{Ivanova_2018}
K.A. Ivanova, S.L. Gavrilyuk, Structure of the hydraulic jump in convergent radial flows, J. Fluid Mech. 860 (2019) 441--464.

\bibitem{Kazakova2019}
M. Kazakova, G.L. Richard, A new model of shoaling and breaking waves: one-dimensional solitary wave on a mild sloping beach, J. Fluid Mech. 862 (2019) 552--591. 

\bibitem{Lannes_March_2015}
D. Lannes, F. Marche, A new class of fully nonlinear and weakly dispersive Green--Naghdi models for efficient 2D simulations, J. Comput. Phys. 282 (2015) 238--68.

\bibitem{Liapidevskii_Chesnokov_2014} 
V.Yu. Liapidevskii, A.A. Chesnokov, Mixing layer under a free surface, J. App. Mech. Tech. Phys. 55 (2014) 299--310. 

\bibitem{Liapidevskii_Gavrilova_2008} 
V.Yu. Liapidevskii, K.N. Gavrilova, Dispersion and blockage effects in the flow over a sill, J. Appl. Mech. Tech. Phys. 49 (2008) 34--45.   

\bibitem{Liapidevskii_2017}
V.Yu. Liapidevskii, V.V. Novotryasov, F.F. Khrapchenkov, I.O. Yaroshchuk, Internal wave bore in the shelf zone of the sea, J. Appl. Mech. Tech. Phys. 58 (2017) 809--818.

\bibitem{L_T_2000} 
V.Yu. Liapidevskii, V.M. Teshukov, Mathematical Models of Propagation of Long Waves in a Non-Homogeneous Fluid, Siberian Branch of Russian Academy of Sciences, Novosibirsk, 2000. [in Russian] 

\bibitem{LeM_G_H_2010}
O. Le Metayer, S. Gavrilyuk, S. Hank, A numerical scheme for the Green--Naghdi model, J. Comput. Phys. 229 (2010) 2034--2045.

\bibitem{Mazaheri_2016}
A. Mazaheri, M. Ricchiuto, H. Nishikawa, A first--order hyperbolic system approach for dispersion, J. Comput. Phys. 321 (2016) 593--605.

\bibitem{Misra_2008}
S.K. Misra, J.T. Kirby, M. Brocchini, F. Veron, M. Thomas, C. Kambhamettu, The mean and turbulent flow structure of a weak hydraulic jump, Phys. Fluids 20 (2008) 035106.

\bibitem{N_T_1990} 
H. Nessyahu, E. Tadmor,  Non-oscillatory central differencing schemes for hyperbolic conservation laws, J. Comput. Phys. 87 (1990)  408--463.

\bibitem{Ovs_1979} 
L.V. Ovsyannikov, Two-layer shallow-water model, J. Appl. Mech. Tech. Phys. 20 (1979) 127--135.

\bibitem{Richard_2012} 
G.L. Richard,  S.L. Gavrilyuk, A new model of roll waves: comparison with Brock's experiments, J. Fluid Mech. 698 (2012) 374–405.

\bibitem{Richard_2013} 
G.L. Richard,  S.L. Gavrilyuk, The classical hydraulic jump in a model of shear shallow-water flows. J. Fluid Mech. 725 (2013) 492–521.

\bibitem{Serre_1953} 
F. Serre, Contribution \`{a} l'\'{e}tude des \'{e}coulements permanents et variables dasn les cannaux, Houille Blanche 8 (1953) 374--388.

\bibitem{Svendsen_2000}
I.A. Svendsen, J. Veeramony, J. Bakunin, J.T. Kirby, The flow in weak turbulent hydraulic jumps, J. Fluid Mech. 418 (2000) 25--57.

\bibitem{Su_Gardner_1969}
C.H. Su, C.S. Gardner, Korteweg -- de Vries equation and generalisations. III. Derivation of the Korteweg -- de Vries equation and Burgers equation, J.  Math. Phys. 10 (1969) 536--539.

\bibitem{Teshukov_2007} 
V.M. Teshukov, Gas-dynamics analogy for vortex free-boundary flows, J. Appl. Mech. Tech. Phys. 48 (2007) 303--309.

\bibitem{Tissier_2012} 
M. Tissier, P. Bonneton, F. Marche, F. Chazel, D. Lannes, A new approach to handle wave breaking in fully non-linear Boussinesq models, Coast. Engng. 67 (2012) 54--66.

\bibitem{Treske_1994}
A. Treske, Undular bores (Favre-waves) in open channels: experimental studies, J. Hydraulic Res. 32 (1994) 355--370.

\bibitem{Townsend_1956} 
A.A. Townsend, The structure of turbulent shear flow, Cambridge University Press, 1956.

\bibitem{Whitham}
G.B. Whitham, Linear and Nonlinear Waves, New-York, John Wiley \& Son, 1999. 

\end{thebibliography}
\end{document}